\documentclass[12pt]{article}

\usepackage{amssymb,amsmath,amsfonts,eurosym,geometry,ulem,graphicx,caption,color,setspace,sectsty,comment,footmisc,caption,natbib,pdflscape,subcaption,array,hyperref,booktabs,floatrow, tikz}
\usepackage{tcolorbox}
\tcbuselibrary{breakable}
\usetikzlibrary{positioning}

\newcolumntype{L}[1]{>{\raggedright\let\newline\\arraybackslash\hspace{0pt}}m{#1}}
\newcolumntype{C}[1]{>{\centering\let\newline\\arraybackslash\hspace{0pt}}m{#1}}
\newcolumntype{R}[1]{>{\raggedleft\let\newline\\arraybackslash\hspace{0pt}}m{#1}}

\begin{document}

\begin{titlepage}
\title{Can Interest-Bearing Positions Solve the Long-Horizon Problem in Prediction Markets? \\
\Large Evidence from Agent-Based Simulations}
\author{Caleb Maresca\thanks{Email: ccm7752@nyu.edu. Department of Economics, New York University. \\
The code for this paper is available at \url{https://github.com/CalebMaresca/prediction-market-experiments}} \\
New York University}
\date{\today}
\maketitle
\begin{abstract}
\noindent Prediction markets suffer from reduced liquidity and price accuracy for long-horizon events due to the opportunity cost of committed capital. Recently, major platforms have introduced interest-bearing positions to mitigate this "long-horizon problem." I evaluate this policy using agent-based simulations with large language model (LLM) traders in a $2 \times 2$ factorial design, varying time horizon (4 days vs. 2 years) and the presence of interest. While long horizons degrade accuracy, the observed pricing bias (0.72 percentage points) is significantly smaller than theoretical and prior empirical estimates. Paying interest eliminates approximately 83\% of the horizon effect on accuracy and more than triples market participation (from 17\% to 62\% of wealth). These findings suggest the long-horizon problem may be overstated in existing literature and that interest-bearing positions are a highly effective intervention, primarily by incentivizing participation rather than correcting bias. \\
\vspace{0in}
\vspace{0in}

\bigskip
\end{abstract}
\setcounter{page}{0}
\thispagestyle{empty}
\end{titlepage}
\pagebreak \newpage

\doublespacing

\section{Introduction} \label{sec:introduction}

Prediction markets harness the price mechanism to aggregate dispersed information into probabilistic forecasts. The intellectual foundation for this approach traces back to \citet{hayek1945use}, who argued that prices efficiently aggregate knowledge dispersed among individuals that no central planner could collect. Building on this insight, \citet{hanson2003combinatorial} advocated for prediction markets as tools for corporate and public decision-making, while \citet{wolfers2004prediction} established prediction markets as a serious topic of academic inquiry, documenting their accuracy across diverse domains. These markets can be of social value: they can improve decision-making in business and government, provide hedging instruments for those with exposure to future events, and generate forecasts that inform public discourse on topics ranging from elections to climate change.

Despite this promise, prediction markets face several practical and theoretical challenges. Markets on political events may create incentives for manipulation \citep{rhode2004historical, rhode2006manipulating}. Thin markets suffer from illiquidity, leading to wide bid-ask spreads and uninformative prices. Regulatory restrictions limit participation in many jurisdictions. This paper focuses on a specific limitation that has received growing attention: the \textit{long-horizon problem}, the tendency for prediction markets to exhibit reduced liquidity and price accuracy when the event will not resolve until far in the future.

The long-horizon problem arises from opportunity costs. A simple example illustrates the mechanism. Consider a trader who believes a contract is mispriced at \$0.50 and that the true probability is 0.60. If the market resolves tomorrow, she expects to earn \$0.10 per contract by buying and holding. But if the market resolves in five years, she still expects to earn only \$0.10, while the same capital invested in a diversified portfolio might earn 50\% or more. In fact, she could earn a higher return for a much lower risk with a bond yielding 4\% yearly. The expected return from prediction market participation is independent of time to resolution, but the returns from outside options grow with time. Consequently, traders with attractive outside options rationally abstain from long-horizon markets, reducing liquidity and potentially biasing prices.

\citet{antweiler2012long} develops an analytical framework formalizing this intuition. He shows that opportunity costs create a participation constraint that reduces liquidity and systematically biases prices in long-horizon markets. His analysis also reveals an important asymmetry: for a contract priced at $p$ dollars, a trader betting the event will occur commits $p$ dollars, while a trader betting against commits $(1-p)$ dollars. At low prices, long positions face smaller opportunity costs than short positions; at high prices, the reverse holds. This asymmetry biases prices toward 50\%, with the distortion growing as time horizons lengthen. \citet{page2013prediction} provide compelling empirical evidence using over 500,000 transactions across 1,787 markets, finding that time until expiration negatively affects price accuracy: markets are ``reasonably well calibrated when time to expiration is relatively short, but prices are significantly biased for events farther in the future.''

To address the long-horizon problem, the two largest prediction markets, Kalshi and Polymarket, have both started offering interest on open market positions.\footnote{\url{https://news.kalshi.com/p/interest-cash-open-positions}}\footnote{\url{https://finance.yahoo.com/news/polymarket-lures-bettors-highest-4-204736721.html}} Polymarket explicitly states that the purpose of this policy is to ``keep long-term pricing accurate,''\footnote{\url{https://docs.polymarket.com/polymarket-learn/trading/holding-rewards}} while Kalshi focuses on the benefits for users, stating ``Interest accrual allows you to hold your positions for a longer period without worrying about running out of funds.''\footnote{\url{https://help.kalshi.com/navigating-the-exchange/your-portfolio/apy-on-kalshi}}

The effectiveness of these policies for improving price accuracy remains unknown. The obvious benefit is that expected returns from prediction market positions now increase with time to resolution, alleviating the long-horizon problem by reducing the gap between prediction market returns and outside option returns. However, there are still reasons why a trader may perceive outside options as superior to long-horizon prediction markets even with interest payments. First, the interest offered by Kalshi and Polymarket is similar to that offered by high-yield savings accounts and T-Bills, which may not be the relevant outside option for many traders. Traders with a higher appetite for risk may prefer to invest in other outside options, such as the stock market. Second, portfolio value interest has a significant disadvantage compared to risk-free alternatives such as T-Bills: traders may lose their entire investment if the market resolves against their position. Furthermore, they likely will not receive a consistent interest payment as a percentage of their initial investment cost, as they would with T-Bills. Rather, the interest is calculated on the current market value of their portfolio, which varies with the price. For example, if a trader bought 1,000 contracts for \$0.50 each, and the market price dropped to \$0.25 immediately thereafter, she would not receive interest on her \$500 investment, but rather on its current value of \$250.

Measuring the effect of portfolio value interest on pricing accuracy using observational data is not currently possible, as it was only instituted by Kalshi in October of 2024 and by Polymarket in September of 2025. Therefore, there are no markets with a horizon of over a year and a half that have resolved since the policies were instituted. It may take a long time to accumulate enough data for an accurate analysis, due not only to the inherent amount of time it takes long-horizon markets to resolve, but also due to their noisy nature. The effect of horizon on mispricing is easily confounded with the fact that markets that are far from resolution often have less information available to help the investor make an accurate prediction. In addition, it might be difficult to find appropriate control markets in which portfolio value interest is absent.

In this paper, I tackle this problem experimentally using Large Language Model (LLM) agents to simulate prediction market trading. I implement a $2 \times 2$ factorial design varying time horizon (short versus long) and portfolio value interest (absent versus present), allowing me to isolate the causal effect of interest-bearing positions on market outcomes. The LLM agents are endowed with heterogeneous risk preferences. Half have high risk tolerance and half have medium risk tolerance, which generates meaningful variation in trading behavior and portfolio allocation decisions.

This approach has several advantages. First, I can remove the confounding effect of information over long time horizons by keeping the amount of information given to the agents constant between the long and short time horizon treatments. Second, I can generate a relatively large dataset with clean controls. Third, I know the true probability of the event occurring. Thus, I can directly compare the market price to the true probability, as opposed to comparing probability to frequency of outcomes, as is required in observational data. Fourth, I can analyze auxiliary information that is unavailable in observational data, such as the agents' predicted probabilities. This is useful because theory predicts that when markets are far from resolution, traders may all know that the market is mispriced, but still choose not to trade due to attractive outside options. Finally, LLM simulations are far cheaper than human experiments (especially as markets require multiple human participants for each observation) allowing for a greater degree of experimentation and larger sample sizes.

This approach also has limitations. The use of LLMs as experimental subjects is still in its infancy, and questions remain about external validity. Recent work finds that LLM-populated markets exhibit ``textbook-rational'' behavior with muted bubble formation compared to human markets \citep{henning2025llmagentsreplicatehuman}, suggesting that LLMs may not reproduce all human behavioral anomalies. However, this limitation is less concerning for the present research, which focuses on whether a specific mechanism design can address a problem that exists even under rational behavior. Additionally, while studies show that LLMs exhibit traits statistically indistinguishable from humans on many behavioral games \citep{mei2024turing}, the mapping from prompt-based ``risk preferences'' to actual trading behavior may not mirror human decision-making in all respects.

The experimental results provide three main findings. First, extending the time horizon from four days to two years reduces price accuracy, trading volume, and prediction market exposure while widening bid-ask spreads, confirming the existence of the long-horizon problem in a controlled setting. Second, the magnitude of the pricing bias attributable to the lengthened time horizon is considerably smaller than both theoretical predictions and prior empirical estimates, roughly 0.72 percentage points compared to the approximately 6 percentage points predicted by extending \citet{antweiler2012long}'s model and the 4.7--10.9 percentage points found empirically by \citet{page2013prediction}. Third, paying interest on prediction market positions eliminates approximately 83\% of the horizon effect on price accuracy and increases market participation by more than 250\%, from 16.9\% to 61.6\%; compared to a short-horizon benchmark of 74.9\%.

The remainder of this paper proceeds as follows. Section \ref{sec:lit} reviews the relevant literature on prediction market accuracy, the long-horizon problem, laboratory prediction market experiments, and the emerging use of large language models in experimental economics. Section \ref{sec:design} describes the experimental design, including the market structure, participant design, and treatment conditions. Section \ref{sec:hypotheses} develops the hypotheses tested in this study. Section \ref{sec:result} presents the main experimental findings. Section \ref{sec:conclusion} concludes with a discussion of the implications for prediction market design and directions for future research.

\section{Literature Review} \label{sec:lit}

\subsection{Prediction Market Accuracy and Information Aggregation}

The theoretical foundation for prediction markets traces to \citet{hayek1945use}, who argued that knowledge is dispersed among individuals and that the price system serves as an efficient mechanism for aggregating this distributed information. The insight that decentralized markets can harness knowledge that no central planner could collect provides the intellectual basis for virtually all prediction market research. The contemporary survey by \citet{wolfers2004prediction} establishes prediction markets as a serious academic topic, documenting that market-generated forecasts typically outperform most moderately sophisticated benchmarks across a range of domains. Their subsequent work, \citet{wolfers2006interpreting}, demonstrates that under log utility, equilibrium prices equal the budget-weighted average of trader beliefs, providing theoretical justification for treating prices as probability estimates.

The Iowa Electronic Markets (IEM) have generated the most rigorous empirical evidence on prediction market accuracy. \citet{berg2008prediction} compared IEM predictions to 964 polls across five U.S. Presidential elections from 1988 to 2004, finding that the market was closer to the eventual outcome 74\% of the time. Election-eve forecasts showed average absolute errors of just 1.33 percentage points, and markets significantly outperformed polls when forecasting more than 100 days in advance. Laboratory evidence from \citet{plott1988rational} demonstrates that double auction trading leads to efficient information aggregation and rational expectations equilibrium under controlled conditions, establishing the experimental foundation for market-based forecasting.

Not all assessments are uniformly positive. \citet{manski2006interpreting} challenges the direct interpretation of prices as probabilities, showing that with risk-neutral price-taking traders and heterogeneous beliefs, prices only partially identify mean beliefs. \citet{wolfers2006interpreting} respond that under reasonable assumptions about risk preferences, prices remain informative about underlying probabilities. This theoretical debate highlights that the mapping from prices to probabilities depends on assumptions about trader preferences and behavior that may not hold in practice. Nevertheless, the empirical record suggests that prediction markets work remarkably well across many domains, whatever the precise theoretical justification.

\subsection{The Long-Horizon Problem}

Despite their demonstrated accuracy over short time horizons, prediction markets consistently suffer from reduced liquidity and price quality when resolution dates are distant. \citet{antweiler2012long} provides the definitive analytical framework for understanding this limitation, identifying the opportunity cost mechanism as the fundamental constraint on long-horizon markets. When traders commit capital to prediction market contracts, that capital cannot be invested in alternative portfolios (such as stocks or bonds) that would otherwise earn a compound return. This creates a participation constraint: traders will only enter the market if their expected gain exceeds this lost opportunity, a threshold that becomes increasingly difficult to meet as the time to resolution lengthens.

Antweiler’s analysis reveals that this opportunity cost does not affect the market uniformly, but rather introduces an important asymmetry rooted in the "money at risk" for different positions. In a standard unit portfolio where a contract pays \$1 upon resolution, a trader taking a "long" position at price $p$ must commit $p$ dollars, while a trader taking a "short" position (betting against the event) must commit $(1-p)$ dollars to the alternate contract. This creates asymmetric capital requirements: at a low market price of $p=0.20$, a long position is significantly "cheaper" to maintain than a short position ($0.20$ vs. $0.80$ at risk).

This asymmetry distorts the equilibrium price because a fixed cent-deviation from beliefs yields a much higher return on the "cheaper" side of the market. For instance, at $p=0.20$, a two-cent mispricing offers an 11.1\% return for a long position but only a 2.5\% return for a short position. This imbalance allows long traders to satisfy their participation constraints more easily than short traders at low prices, creating upward pressure that results in a systematic long bias. Conversely, at high prices, short positions become more capital-efficient, leading to a short bias. Equilibrium is only achieved when the market price deviates from the mean trader belief enough to equalize these asymmetric participation rates, distorting predictions toward the 50\% mark.

Empirical evidence supports these theoretical predictions. \citet{page2013prediction} analyze over 500,000 transactions across 1,787 markets on Intrade, finding that time until expiration negatively affects price accuracy. High-likelihood events are systematically underpriced and low-likelihood events are overpriced, with markets reasonably well calibrated when time to expiration is relatively short but significantly biased for events farther in the future. Page and Clemen invoke a different theoretical reason for the bias. Rather than comparing the prediction market to outside options as in \citet{antweiler2012long}, they show that the discounting of future payoffs implies that traders whose beliefs are near the market price abstain from participation, forming a "no-trade region" where neither buying nor selling is utility-maximizing. Because the cost of time discounting is proportional to the capital committed, this abstention is more pronounced for high-priced ``favourite" contracts, which require a larger upfront investment. This asymmetric reduction in trade creates an imbalance where demand for favourites decreases more than supply when prices are high, while supply for low-priced ``longshots" decreases more than demand. The net result of this "disappearance of trade" is a systematic longshot bias where equilibrium prices are distorted toward $0.50$ as the time until expiration increases.

\subsection{Laboratory Prediction Market Experiments}

The transition from theoretical models to empirical validation began with laboratory experiments designed to test whether the price mechanism could truly aggregate dispersed information. This tradition was pioneered by \citet{plott1982efficiency} and \citet{plott1988rational}, who utilized double auction markets to observe how information is shared among traders. While \citet{plott1982efficiency} demonstrated that markets could disseminate information from "insiders" to uninformed traders, the 1988 study provided the crucial evidence for information aggregation—showing that a market can pool diverse, noisy signals to identify a state that no individual trader knows with certainty. These early studies established that, under controlled conditions, prices in double auctions frequently converge to rational expectations equilibria.

Subsequent laboratory research has identified the behavioral and institutional limits of this aggregation. \citet{forsythe1990information} clarified that information aggregation is more likely when traders have repeated experience with the market institution and common knowledge of the dividend structure. Furthermore, work by \citet{forsythe1992anatomy} and \citet{oliven2004suckers} analyzed laboratory and field equivalents of the Iowa Electronic Markets, finding that pricing efficiency often rests on a small subset of "marginal traders" who correct the biases of less active participants. These findings emphasize that while markets are theoretically robust, their real-world performance is contingent on specific institutional and participant characteristics, such as the speed of convergence and the distribution of private information.

Providing a comprehensive synthesis of this literature, \citet{deck2013prediction} offer a broad review of the field's evolution. Their work contextualizes the "wisdom of crowds" within the framework of markets as information processors, distilling diverse beliefs into singular prices. The review documents the expansion of prediction markets from foundational studies into varied domains, including corporate sales forecasting, product launch tracking, and scientific discovery. Ultimately, \citet{deck2013prediction} conclude that the reliability of market-generated forecasts is not universal but depends on environmental factors such as market thickness, the simplicity of payoff structures, and the intensity of incentives to manipulate. This synthesis identifies that maintaining robust, ongoing participation remains a critical challenge for ensuring market efficiency in complex or long-term settings.

\subsection{Large Language Models in Experimental Economics}

A growing literature explores the use of large language models as simulated economic agents. \citet{horton2023large} coins the term ``Homo Silicus'' and establishes that LLMs can be given endowments, preferences, and information like traditional economic agents, replicating classic experiments with results qualitatively similar to human subjects. This approach offers a method for piloting experiments before costly human studies. This benefit is especially useful for market experiments that require a large number of subjects.

Importantly, Horton shows that while LLMs have likely ingested the results of many experiments in their training corpus, they do not overfit to past results. Instead, they can generalize to novel scenarios. Horton demonstrates this robustness by varying original experiments through linguistic translations, context changes—such as substituting snow shovels for artisan bread—and even adversarial prompts designed to invert established behavioral relationships. Despite these perturbations, the models maintain consistent patterns, suggesting they have internalized a generalized "implicit social information" drawn from both scholarly literature and the vast records of human interactions in their training data.

The sensitivity of LLM performance to experimental design has also become a key focus. \citet{wang2025experimentaleconomicsmeetslarge} demonstrate that while assigned personas (such as occupation) significantly influence LLM preferences, they do not substantially impact economic rationality. Furthermore, they find that the choice of dialogue structure is critical; using single-turn dialogues significantly decreases the rationality scores of smaller models compared to multi-turn interactions.

Evidence on LLM rationality and behavioral alignment is mixed but encouraging for certain applications. \citet{chen2023emergence} demonstrate that GPT achieves higher rationality scores than human subjects across risk, time, social, and food preference domains. \citet{mei2024turing} provide the most comprehensive behavioral comparison, showing that GPT-4 exhibits traits statistically indistinguishable from random humans across over 50 countries on classic behavioral games, including the Dictator Game, Ultimatum Game, Trust Game, and Prisoner's Dilemma. When deviating from human norms, LLMs tend to be more cooperative and altruistic rather than less.

For market simulations specifically, \citet{lopez2025can} demonstrates that LLMs can act as consistent value investors, momentum traders, or market makers, with market dynamics exhibiting realistic features including price discovery and strategic liquidity provision. However, \citet{henning2025llmagentsreplicatehuman} find that LLM-populated markets exhibit 'textbook-rational behavior,' pricing assets near fundamental value with muted bubble formation compared to experimental markets with human participants. This finding suggests that LLM agents may be too rational to generate the behavioral anomalies characteristic of human markets, but may be well-suited to studying market design features where rational-agent predictions are theoretically interesting. Recent work by \citet{delriochanona2025generativeaiagentsbehave} nuances this finding, showing that in dynamic market experiments, LLMs do not strictly follow rational expectations but instead display bounded rationality similar to human participants. They emphasize that providing a context window or "memory" of at least three previous time steps is necessary for LLMs to replicate broad market trends.

Concerns regarding whether LLMs are more rational than human participants are less concerning for the present research, which focuses on whether a specific mechanism design (interest-bearing positions) can address a problem that exists even under rational behavior. The opportunity cost mechanism identified by \citet{antweiler2012long} operates through rational capital allocation decisions, not behavioral biases. If anything, the rationality of LLM agents provides a clean test of the theoretical mechanism: if even rational agents reduce participation in long-horizon markets due to opportunity costs, and if paying interest mitigates this effect, the results speak directly to the policy question without confounding from behavioral factors.

\section{Experimental Design} \label{sec:design}

\subsection{Experiment Structure} \label{sec:design:sub:experiment-structure}

The experiment implements a binary prediction market for a single event with known resolution date. The event occurs with probability $p^*$, which is unknown to the agents. Markets operate as double auctions where participants can trade contracts that pay \$1 if the event occurs (``Yes'' contracts) and \$0 otherwise, or contracts that pay \$1 if the event does not occur (``No'' contracts) and \$0 otherwise. In addition, the agents have two available outside options: a risk-free asset and a risky asset.

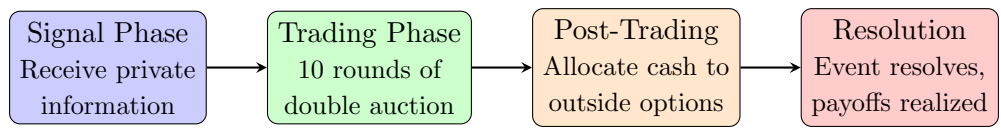
\begin{figure}[htbp]
\centering
\begin{tikzpicture}[node distance=0.8cm, phase/.style={rectangle, draw, rounded corners, minimum width=2.5cm, minimum height=1cm, align=center, font=\small}, arrow/.style={->, thick, >=stealth}]
\node[phase, fill=blue!20] (signal) {Signal Phase\\{\footnotesize Receive private}\\{\footnotesize information}};
\node[phase, fill=green!20, right=of signal] (trading) {Trading Phase\\{\footnotesize 10 rounds of}\\{\footnotesize double auction}};
\node[phase, fill=orange!20, right=of trading] (post) {Post-Trading\\{\footnotesize Allocate cash to}\\{\footnotesize outside options}};
\node[phase, fill=red!20, right=of post] (resolution) {Resolution\\{\footnotesize Event resolves,}\\{\footnotesize payoffs realized}};
\draw[arrow] (signal) -- (trading);
\draw[arrow] (trading) -- (post);
\draw[arrow] (post) -- (resolution);
\end{tikzpicture}
\caption{Experiment Timeline}
\label{fig:timeline}

\floatfoot{\textit{Notes:} The experiment proceeds in four stages. In the signal phase, agents receive private information (200 Bernoulli draws). During the trading phase, agents participate in 10 rounds of double auction trading. In the post-trading phase, agents allocate remaining cash between risk-free and risky outside options. Finally, the event resolves after horizon $\tau$ (4 days or 2 years depending on treatment) and payoffs are realized.}

\end{figure}

Figure \ref{fig:timeline} illustrates the experiment timeline. Each experimental session proceeds in three phases: the signal phase, the trading phase, and the post-trading phase.

In the signal phase at the start of the experiment, each agent receives private information about the true event probability $p^*$. Specifically, each agent observes $S = 200$ independent draws from a Bernoulli distribution with the true probability $p^*$. Agents see how many of these draws were ``successes,'' providing them with a noisy but informative signal about the underlying probability. With 200 draws, agents can form reasonably precise posterior beliefs while retaining meaningful uncertainty that motivates trade.

The true probability is set to either $p^* = 0.05$ or $p^* = 0.95$, representing events at the extremes of the probability distribution where the long-short asymmetry identified by \citet{antweiler2012long} predicts the largest price distortions. However, the agents are not informed of the experimental distribution of true probabilities and are instead told ``The event has a true probability of occurring that is fixed and lies in the interval [0.0, 1.0]. This true probability is unknown to all participants." Examining the reasoning summaries provided with responses from the OpenAI API, I find that agents interpret this as either a uniform prior or a Jeffreys prior ($Beta(1/2,1/2)$).

In the trading phase, agents participate in $R = 10$ discrete trading rounds in a double auction market for the ``Yes'' and ``No'' contracts. Agents trade using their full cash endowment during this phase. Following \citet{lopez2025can}, I use discrete trading rounds rather than continuous real-time order processing because LLM latency constraints make real-time processing infeasible. Within each round, agent orders are collected, then randomized to simulate concurrent order arrival while maintaining fairness, and finally processed according to standard price-time priority rules.

The market operates on both ``Yes'' and ``No'' sides simultaneously through a unified order book with an important structural feature: a bid of $p$ to buy a ``Yes'' contract automatically appears as an ask to sell a ``No'' contract at $(1-p)$, and vice versa. This duality arises naturally from the complementary payoff structure: if I am willing to buy a Yes contract for \$0.60, I am implicitly offering to be the counterparty to someone who wants to buy a No contract for \$0.40, since together we would pay exactly \$1.00 to fund the eventual payout. Such a structure is used in both Kalshi and Polymarket.

Short selling either contract type is not permitted, and is not necessary due to the complementary nature of the two contract types. Additionally, orders persist across trading rounds unless explicitly canceled by agents. This persistence enables price discovery and reduces the need for agents to constantly re-submit orders while allowing them to adapt their strategies based on market developments.

In the post-trading phase, after the final trading round, agents make a single portfolio allocation decision: they choose what fraction of their remaining cash to invest in the risky outside option versus the risk-free outside option. In treatments with portfolio value interest, agents also receive interest on the market value of their prediction market positions, calculated using the price at the end of the final trading round. The event then resolves, outside investment returns are realized, and final payoffs are determined.

While this experimental design shares the fundamental double-auction structure of classic laboratory studies, it introduces several key departures to address the specific requirements of the long-horizon problem. Unlike the foundational work of \citet{plott1982efficiency} and \citet{plott1988rational}, which typically utilized human subjects and discrete state spaces (e.g., states X, Y, and Z), this design employs LLM agents and a continuous probability space $p^* \in [0, 1]$. Whereas classic designs focused on convergence toward a single "correct" discrete state, using a continuous truth allows for a more granular measurement of price accuracy and systematic bias at different time horizons through mean absolute error. Furthermore, while classic experiments relied on continuous real-time auctions, this paper implements discrete trading rounds to accommodate LLM latency and ensure a level playing field for order execution. Lastly, the inclusion of explicit risky and risk-free outside options, an element absent from traditional designs, is essential for simulating the capital opportunity costs that define the long-horizon problem.

\subsection{Participant Design}

The experimental markets consist of $N=10$ LLM agents implemented using the OpenAI LLM GPT-5.2. Five agents are assigned high risk tolerance, and five are assigned medium risk tolerance through natural language persona descriptions in their system prompts. Pilot simulations also included agents with low risk tolerance, but they rarely participated in the prediction market, preferring to allocate most or all of their capital to the outside options, and were therefore removed from the experiment to focus on agents who actively engage with the prediction market. 

This approach follows the emerging methodological literature on LLM experiments. \citet{horton2023large} establishes that LLMs can be induced to ``play different agents via prompts," noting that this agent programming ``is not unlike the experimental economics practice of giving an experimental subject a card that says their marginal cost of producing a widget is 15 tokens.'' More recent work by \citet{wang2025experimentaleconomicsmeetslarge} finds that assigned personas such as occupation or risk profile ``significantly affect preferences but not rationality,'' suggesting that persona-based heterogeneity operates similarly to how naturally occurring preference heterogeneity functions in human experiments.

I chose not to induce preferences through explicit utility functions for two reasons. First, specifying a particular functional form would limit the experiment to testing whether that specific model responds to opportunity costs, rather than testing the mechanism more generally. Second, this approach better mirrors the conditions of actual prediction markets: when human traders participate, they bring naturally heterogeneous risk preferences that we do not observe or control. Just as experimenters conducting laboratory experiments with human subjects cannot specify participants' utility functions and must work with whatever preferences people bring into the lab, I allow the LLM agents to interpret their assigned risk tolerances and translate them into trading behavior. This creates meaningful variation in trading and portfolio allocation decisions without imposing assumptions about the precise shape of utility that might not hold for human traders.

The LLM agents operate through a structured prompt-response loop. At the start of the experiment, agents receive a system prompt explaining the experiment structure, their initial endowment of $W_0 = \$10{,}000$, the number of participants, their objective (to maximize final wealth), and their risk profile. They also receive their private signal: the number of successes observed in their 200 Bernoulli draws.

Each trading round, agents are prompted once to provide their trading decisions. Each prompt consists of the current round number, updates from the previous trading round (e.g., orders created, orders filled), the market state for both Yes and No contracts (including best bid and ask prices and the full order book), and their current financial position. Agents respond by providing a probability estimate for the event, a list of orders they wish to create, edit, or cancel, and explanations for their decisions.

After the final trading round, agents receive a separate prompt for their post-trading investment decision. They are informed of the outcome of trading, their current cash position and contract holdings, and the parameters of the outside investment options. They then specify what fraction of their remaining cash to allocate to the risky asset, with the remainder going to the risk-free asset.

For each agent, the entire conversation history of all prompts and responses is saved. At each decision point, the agent receives not only the current prompt but the full history of prior interactions, enabling agents to adapt their strategies based on observed market behavior and their own performance. This architecture mirrors human memory in laboratory experiments, where participants can recall their previous actions and market observations.

The information structure follows standard conventions in experimental economics. The experiment rules, number of participants, contract payoffs, and outside option parameters are common knowledge. All agents receive identical information about these aspects in their system prompts. Private signals (the number of Bernoulli successes each agent observes) remain private information. Agents know only their own signal. The order book and transaction history become public information as trading proceeds, disseminated to all agents in their round prompts. This structure mirrors the classic \citet{plott1988rational} design, where traders had private information about the state but observed public market prices.

Appendix \ref{sec:appendixb} presents a partial transcript from one experimental session, including the full system prompt, example trading round prompt and response, and example post-trading investment decision prompt and response.

\subsection{Treatments}

The experimental design consists of four cells organized as a $2 \times 2$ factorial design, varying time horizon and interest on positions. Table \ref{tab:factorial} summarizes the treatment structure.

\begin{table}[ht]
\centering
\caption{Experimental Design: $2 \times 2$ Factorial Structure}
\label{tab:factorial}
\begin{tabular}{lcc}
\toprule
& \textbf{Short Horizon (4 days)} & \textbf{Long Horizon (2 years)} \\
\midrule
\textbf{Without Interest} & Cell 1 & Cell 2 \\
 & $r_f = 0.044\%$, $\mu_r = 0.11\%$, $\sigma_r = 1.7\%$ & $r_f = 8\%$, $\mu_r = 20\%$, $\sigma_r = 22.63\%$ \\
& No portfolio interest & No portfolio interest \\
\midrule
\textbf{With Interest} & Cell 3 & Cell 4 \\
 & $r_f = 0.044\%$, $\mu_r = 0.11\%$, $\sigma_r = 1.7\%$ & $r_f = 8\%$, $\mu_r = 20\%$, $\sigma_r = 22.63\%$ \\
& Portfolio interest at $r_f$ & Portfolio interest at $r_f$ \\
\bottomrule
\end{tabular}
\floatfoot{\textit{Notes:} Each cell contains 100 independent simulation sessions with 10 agents each. Parameters $r_f$, $\mu_r$, and $\sigma_r$ denote the risk-free return, expected risky return, and risky return standard deviation over the relevant horizon. Short-horizon parameters are scaled from annual rates over 4 days; long-horizon parameters are compounded over 2 years.}
\end{table}

The time horizon treatment manipulates the economic significance of time while holding constant the number of discrete trading rounds and the quality of information signals. In the short-horizon treatment, the event resolves in four days, making the outside investment options economically negligible. In the long-horizon treatment, the event resolves in two years, making the outside options economically meaningful.

The outside option parameters are calibrated to approximate realistic financial market conditions. The annualized risk-free rate of 4\% reflects recent yields on short-term Treasury securities, as well as the interest rates offered on position in Kalshi and Polymarket\footnote{As of this writing, Kalshi offers a 3.25\% interest rate and PolyMarket offers a 4\% rate.}. The risky asset parameters (10\% expected return and 16\% annual volatility) are calibrated to long-run historical performance of the S\&P 500, which has delivered average annual returns of approximately 10\% with a standard deviation of roughly 15--20\% over the past century. These parameters fall within the range of subjective return expectations documented in survey evidence. \citet{greenwood2014expectations} find that investor expectations for stock market returns average around 6--10\% annually, while \citet{adam2021survey} show that subjective expected returns in survey data are on average close to unbiased relative to subsequent realized returns.

In both treatments, agents make a single post-trading allocation decision between the risk-free asset and the risky asset. The parameters are scaled to the appropriate horizon using standard financial compounding conventions. For the short horizon of four days (approximately $4/365 \approx 0.011$ years), the risk-free return is $r_f = (1.04)^{4/365} - 1 \approx 0.044\%$. The expected risky return scales as $\mu_r = 0.10 \times (4/365) \approx 0.11\%$, and volatility scales with the square root of time as $\sigma_r = 0.16 \times \sqrt{4/365} \approx 1.7\%$. For the long horizon of two years, the compounded parameters are $r_f = (1.04)^2 - 1 = 8.16\% \approx 8\%$, $\mu_r = (1.10)^2 - 1 = 21\% \approx 20\%$, and $\sigma_r = 0.16 \times \sqrt{2} \approx 22.63\%$. This calibration ensures that in the long-horizon treatment, agents face a meaningful opportunity cost: capital held in prediction market positions could instead earn returns in the outside options. In the short-horizon treatment, this opportunity cost is negligible.

The interest treatment varies whether agents earn interest on capital invested in prediction market positions. In the without-interest treatment, prediction market investments earn no interest, and agents forgo the risk-free rate on any capital committed to prediction market contracts. This represents the traditional opportunity cost structure where capital locked in prediction markets cannot earn returns elsewhere. In the with-interest treatment, agents earn interest at the risk-free rate on the market value of their prediction market portfolio, calculated using prices from the final trading round. This models the policy innovation recently adopted by platforms such as Kalshi and Polymarket. Importantly, this differs from the risk-free asset: while the interest rate is identical, the portfolio value on which interest is calculated fluctuates with market prices, creating price risk that is absent from the risk-free outside option.

This $2 \times 2$ design enables direct pairwise comparisons that test the core hypotheses. Comparing Cell 2 to Cell 1 identifies the baseline time-horizon effect in the absence of interest. Comparing Cell 4 to Cell 2 isolates the effect of introducing portfolio value interest in long-horizon markets. Comparing Cell 4 to Cell 3 reveals whether a time-horizon effect persists even when interest is paid on positions. These pairwise comparisons provide clean answers to the central policy question: does paying interest on prediction market positions mitigate the long-horizon problem?

Each of the four experimental cells contains 100 independent simulation sessions, with 50 sessions using $p^* = 0.05$ and 50 sessions using $p^* = 0.95$. This balanced design at extreme probabilities serves two purposes. First, the regions very close to the boundaries ${0,1}$ are where the opportunity cost mechanism predicts the largest price distortions: at $p^* = 0.05$, long positions require only \$0.05 at risk while short positions require \$0.95, creating near maximum asymmetry in participation constraints, increasing statistical power. Second, using symmetric extreme values allows pooling across sessions when measuring absolute pricing errors, since the theoretical bias magnitude is symmetric around 0.50. The total of 400 independent simulation runs provides sufficient statistical power to detect economically meaningful differences in price accuracy and trading volume across treatment conditions.

\section{Hypotheses} \label{sec:hypotheses}

My experiment is designed to test two primary hypotheses about how interest-bearing positions affect prediction market dynamics across time horizons.

\subsection{Hypothesis 1: The Long-Horizon Effect} \label{sec:hypotheses:subsec:H1}

The first hypothesis concerns the baseline time-horizon effect in the absence of portfolio value interest. The opportunity cost framework developed by \citet{antweiler2012long} predicts that when capital committed to prediction markets earns no return, informed traders face an opportunity cost that grows with the time until resolution. Over short horizons, this opportunity cost is negligible: an agent holding a position for four days forgoes minimal returns from outside options. Over longer horizons, however, the foregone returns are considerably larger. In the two-year horizon treatment, the risk-free outside option offers an 8\% return, while the risky option offers an expected 20\% return. 

To generate a quantitative benchmark for the expected magnitude of this effect, I extend Antweiler's model to incorporate the specific features of my experimental design, including heterogeneous risk preferences and the availability of a risky outside option (see Appendix \ref{sec:appendixa} for details). The extended model predicts that at a two-year horizon with a true probability of 0.05, the equilibrium price would be approximately 0.11, a bias of roughly 6 percentage points toward 0.50. Likewise, a true probability of 0.95 would also result in a bias of roughly 6 percentage points. Evaluating the extended model at a four-day horizon results in a bias of only 0.028 percentage points, with this negligible bias arising entirely because agents' posterior mean beliefs are slightly shrunk toward 0.5 by the uniform prior.

Empirical evidence from \citet{page2013prediction} provides additional context for calibrating expectations. Using over 500,000 transactions from Intrade, they find that markets with more than 100 days to expiration exhibit systematic biases of 4.7--10.9 percentage points when evaluated at prices of 0.20 and 0.80. However, their parametric model, when extrapolated to a two-year horizon and evaluated at true probabilities of 0.05 and 0.95, produces implausibly large biases exceeding 30 percentage points, suggesting that the relationship between time horizon and bias may be nonlinear or that other mechanisms attenuate the effect at very long horizons.

This leads to the first hypothesis:

\textit{H1: In the absence of interest on positions, long-horizon markets will exhibit greater mispricing, lower trading activity, lower prediction market exposure, and wider bid-ask spreads compared to short-horizon markets.}

I measure mispricing as the mean absolute error between market prices and true probabilities, which captures the market's accuracy relative to ground truth. Following the convention used by major prediction market platforms such as Kalshi, I use the last trade price as the primary price measure; robustness analyses using the bid-ask midpoint (as employed by Polymarket) are reported in Appendix \ref{sec:appendixc}. I also examine the mean absolute error between market prices and agents' average probability beliefs, which captures whether prices reflect the information that traders actually possess. The distinction matters theoretically. If long-horizon markets exhibit higher price-truth errors but similar price-belief errors, this would suggest that agents trade on their beliefs but those beliefs are somehow less accurate in long-horizon settings; a pattern inconsistent with the experimental design, which holds information quality constant. If instead long-horizon markets exhibit higher errors on both measures, this indicates that agents fail to move prices toward their beliefs, consistent with the opportunity cost mechanism discouraging informed trading.

I examine three additional outcome measures. First, trading volume captures overall market activity: the opportunity cost theory predicts that agents will trade less actively when time horizons are long and foregone returns from outside options are large. Second, prediction market exposure, defined as the value of an agent's prediction market contracts as a share of total wealth, captures the extent to which agents commit capital to the market. Theory predicts lower exposure in long-horizon markets as agents prefer to keep capital available for outside investment options. Third, bid-ask spreads measure market liquidity: with fewer participants willing to commit capital, order books should thin out, leading to wider spreads in long-horizon markets.

To test H1, I compare Cell 2 (long horizon, no interest) against Cell 1 (short horizon, no interest) using OLS regressions with treatment indicator variables. The coefficient on the long-horizon indicator captures the effect of extending the time horizon in the absence of interest. For price accuracy measures and bid-ask spreads, I test whether Cell 2 exhibits significantly higher values than Cell 1. For trading volume and prediction market exposure, I test whether Cell 2 exhibits significantly lower values than Cell 1. These directional predictions follow from the opportunity cost theory. Non-parametric Mann-Whitney U tests are reported in Appendix \ref{sec:appendixc} as robustness checks.

\subsection{Hypothesis 2: The Effect of Portfolio Value Interest}

The second hypothesis concerns the effectiveness of portfolio value interest as a policy intervention. If the opportunity cost of foregone risk-free interest is the binding constraint that drives traders away from long-horizon markets, then paying interest on prediction market positions should mitigate or eliminate this effect. The key empirical question is the magnitude of this mitigation: does paying interest fully eliminate the time-horizon penalty, substantially reduce it, or have only modest effects?

\textit{H2: Paying interest on prediction market positions will reduce the gap in price accuracy, trading activity, prediction market exposure, and bid-ask spreads between short-horizon and long-horizon markets.}

Testing H2 requires two complementary comparisons. First, I compare Cell 4 (long horizon, with interest) against Cell 2 (long horizon, no interest) to isolate the direct effect of introducing portfolio value interest in long-horizon markets. If interest payments mitigate the long-horizon problem, Cell 4 should exhibit lower pricing errors, higher trading volume, higher prediction market exposure, and narrower bid-ask spreads than Cell 2. Second, I compare Cell 4 against Cell 3 (short horizon, with interest) to determine whether a residual time-horizon effect persists even when interest is paid. If paying interest fully eliminates the opportunity cost mechanism, there should be no significant difference between Cell 4 and Cell 3 on any outcome measure. If Cell 4 continues to exhibit higher pricing errors, lower trading activity, lower prediction market exposure, and wider bid-ask spreads than Cell 3, this suggests that opportunity costs beyond the risk-free rate, such as the opportunity cost of risk capital that could earn higher returns in the risky outside option, continue to deter participation in long-horizon markets even when platforms pay interest on positions.

I test H2 using OLS regressions that include an interaction term between the long-horizon and interest treatment indicators. In the regression specification $Y_i = \beta_0 + \beta_1 \cdot \text{LongHorizon}_i + \beta_2 \cdot \text{Interest}_i + \beta_3 \cdot (\text{LongHorizon}_i \times \text{Interest}_i) + \varepsilon_i$, the interaction coefficient $\beta_3$ directly captures the extent to which interest payments mitigate the horizon effect. A negative $\beta_3$ for pricing errors and bid-ask spreads (or a positive $\beta_3$ for trading volume and prediction market exposure) indicates that interest reduces the gap between short-horizon and long-horizon market outcomes. The sum $\beta_1 + \beta_3$ measures the residual horizon effect that persists even with interest payments: if this sum is close to zero and statistically insignificant, interest payments fully eliminate the horizon penalty; if it remains significant, some horizon effect persists. Non-parametric Mann-Whitney U tests for the pairwise comparisons are reported in Appendix \ref{sec:appendixc} as robustness checks.

The combination of these tests provides a complete picture of the policy intervention's effectiveness. Confirmation of H1 paired with full mitigation under H2 would validate the interest-bearing policy as an effective solution to the long-horizon problem. Confirmation of H1 paired with only partial mitigation under H2 would suggest that additional mechanisms are required to achieve parity between short-horizon and long-horizon markets.

\section{Results} \label{sec:result}

This section presents the main experimental findings. I begin with an overview of the results, then examine treatment effects on each outcome measure using OLS regressions, and conclude with an analysis of market dynamics over trading rounds. Non-parametric Mann-Whitney U tests and robustness analyses using alternative price measures are reported in Appendix \ref{sec:appendixc}.

\subsection{Overview}

The experimental results confirm the existence of the long-horizon problem in a controlled setting: extending the time horizon reduces price accuracy, all else equal. However, while the effect is statistically significant, its magnitude is considerably smaller than both the theoretical predictions of \citet{antweiler2012long} and the empirical estimates from \citet{page2013prediction}. Introducing portfolio value interest brings long-horizon market performance close to short-horizon benchmarks.

\begin{figure}[htbp]
\centering
\includegraphics[width=1.0\textwidth]{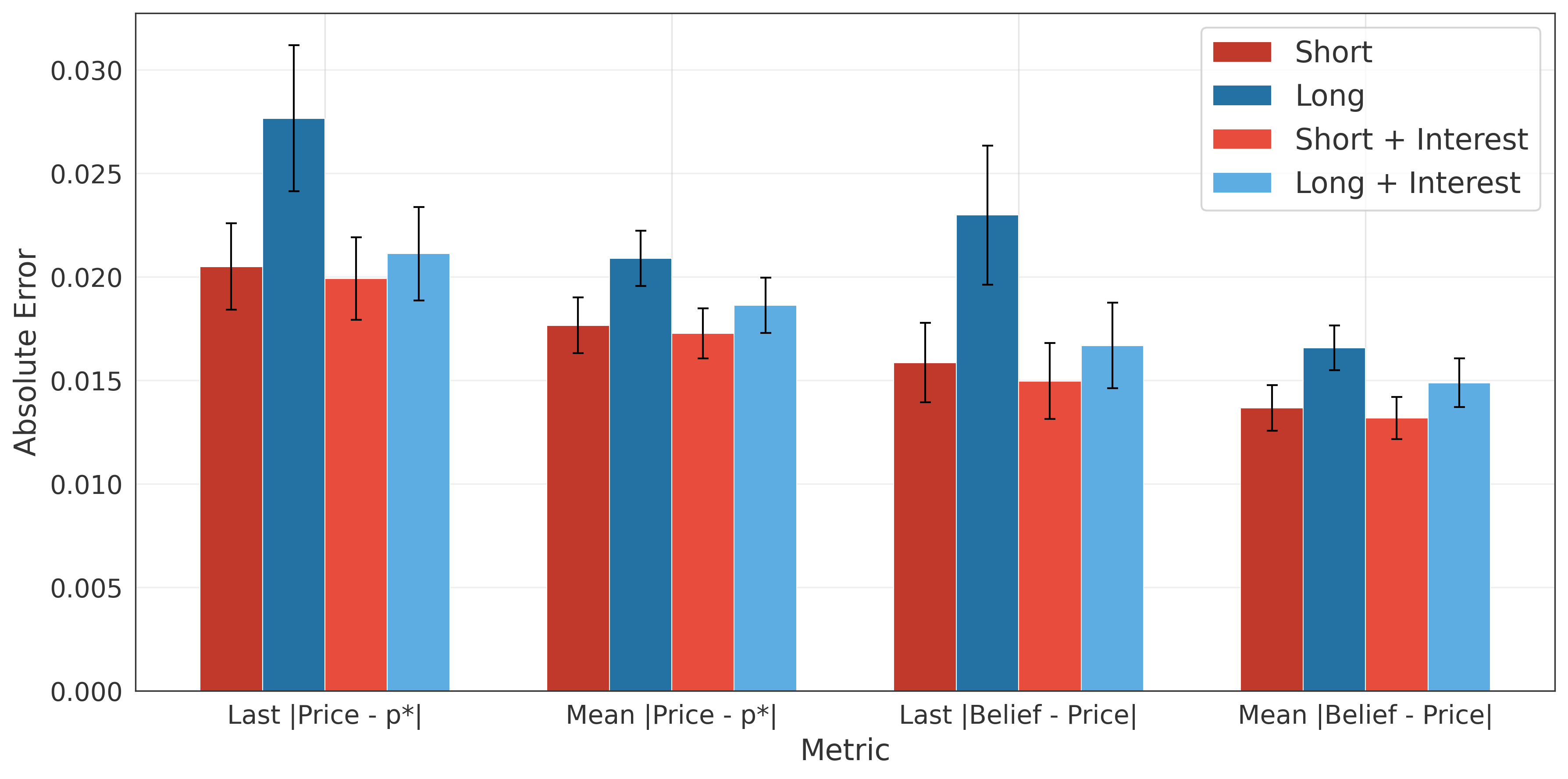}
\caption{Price Accuracy Measures by Treatment}
\label{fig:summary_bars}
\floatfoot{\textit{Notes:} Mean absolute error between market prices and true probabilities (left two groups) and between market prices and average agent beliefs (right two groups). ``Last'' refers to the final trading round price; ``Mean'' refers to the average across all rounds. Error bars show 95\% confidence intervals. $N = 100$ sessions per treatment.}
\end{figure}

Figure \ref{fig:summary_bars} provides a visual summary of the main price accuracy measures across all four treatment cells. Long-horizon markets without interest exhibit the highest pricing errors, while the three other treatment cells cluster together with lower errors. Notably, long-horizon markets with interest show price accuracy that falls well within the 95\% confidence intervals of short-horizon markets, suggesting that interest payments largely eliminate the horizon effect on price accuracy.

\begin{table}[htbp]
\centering
\caption{Summary Statistics by Treatment Cell}
\label{tab:summary_stats}
\begin{tabular}{lcccc}
\toprule
& \multicolumn{2}{c}{\textbf{Without Interest}} & \multicolumn{2}{c}{\textbf{With Interest}} \\
\cmidrule(lr){2-3} \cmidrule(lr){4-5}
& Short (Cell 1) & Long (Cell 2) & Short (Cell 3) & Long (Cell 4) \\
\midrule
Last $|\text{Price} - p^*|$ & 0.020 & 0.028 & 0.020 & 0.021 \\
Mean $|\text{Price} - p^*|$ & 0.018 & 0.021 & 0.017 & 0.019 \\
Last $|\text{Price} - \text{Belief}|$ & 0.016 & 0.023 & 0.015 & 0.017 \\
Mean $|\text{Price} - \text{Belief}|$ & 0.014 & 0.017 & 0.013 & 0.015 \\
Last PM Exposure (\%) & 74.9 & 16.9 & 74.4 & 61.6 \\
Mean PM Exposure (\%) & 59.0 & 22.1 & 59.1 & 56.2 \\
Last Bid-Ask Spread & 0.013 & 0.027 & 0.011 & 0.023 \\
Total Volume (\$) & 135,239 & 120,057 & 143,590 & 157,395 \\
Mean Bid-Ask Spread & 0.012 & 0.017 & 0.011 & 0.014 \\
\midrule
$N$ (sessions) & 100 & 100 & 100 & 100 \\
\bottomrule
\end{tabular}
\end{table}

Table \ref{tab:summary_stats} presents summary statistics for the primary outcome variables across all four experimental cells. Cell 2 (long horizon, no interest) exhibits the highest pricing errors, lowest trading volume, lowest prediction market exposure, and widest bid-ask spreads, while the three other cells cluster together with better performance.

To test hypotheses H1 and H2 and obtain interpretable effect size estimates, I estimate OLS regressions of the form:

\begin{equation}
Y_i = \beta_0 + \beta_1 \cdot \text{LongHorizon}_i + \beta_2 \cdot \text{Interest}_i + \beta_3 \cdot (\text{LongHorizon}_i \times \text{Interest}_i) + \varepsilon_i
\end{equation}

where $Y_i$ is the outcome for session $i$, $\text{LongHorizon}_i$ is an indicator for the two-year horizon treatment, and $\text{Interest}_i$ is an indicator for the portfolio value interest treatment. The coefficient $\beta_1$ captures the long-horizon effect in the absence of interest, directly testing H1. The coefficient $\beta_3$ captures whether interest payments mitigate the horizon effect, testing H2. The intercept $\beta_0$ represents the baseline (short horizon, no interest).

\subsection{Price Accuracy}

Table \ref{tab:ols_price} presents OLS regression results for the price accuracy measures. I examine both the mean absolute error between market prices and true probabilities (Price-Truth MAE) and the mean absolute error between market prices and agents' average beliefs (Price-Belief MAE), using both the the final round price and average across all trading rounds. The final price is preferred in prediction market experiments, as it can take time for the private information of the participants to be fully reflected in the market \citep{deck2013prediction}.

\begin{table}[htbp]
\centering
\caption{OLS Regression Results: Price Accuracy}
\label{tab:ols_price}
\begin{tabular*}{\textwidth}{@{\extracolsep{\fill}}lcccc}
\toprule
& \multicolumn{2}{c}{\textbf{Price-Truth MAE}} & \multicolumn{2}{c}{\textbf{Price-Belief MAE}} \\
\cmidrule(lr){2-3} \cmidrule(lr){4-5}
& Last Round & Mean & Last Round & Mean \\
\midrule
Intercept & 0.0205*** & 0.0177*** & 0.0159*** & 0.0137*** \\
& (0.001) & (0.001) & (0.001) & (0.001) \\
Long Horizon & 0.0072*** & 0.0032*** & 0.0071*** & 0.0029*** \\
& (0.002) & (0.001) & (0.002) & (0.001) \\
Interest & $-$0.0006 & $-$0.0004 & $-$0.0009 & $-$0.0005 \\
& (0.002) & (0.001) & (0.002) & (0.001) \\
Long $\times$ Interest & $-$0.0060** & $-$0.0019 & $-$0.0054** & $-$0.0012 \\
& (0.003) & (0.001) & (0.002) & (0.001) \\
\midrule
$R^2$ & 0.057 & 0.038 & 0.064 & 0.041 \\
$N$ & 400 & 400 & 400 & 400 \\
\bottomrule
\end{tabular*}
\parbox{\textwidth}{\footnotesize\textit{Notes:} OLS regressions of price accuracy measures on treatment indicators. Standard errors in parentheses. * $p<0.10$, ** $p<0.05$, *** $p<0.01$.}
\end{table}

The results confirm H1: the long-horizon effect is statistically significant across all price accuracy measures. In the absence of interest, extending the time horizon from four days to two years increases mean absolute error relative to true probabilities by 0.72 percentage points ($p < 0.001$) for the final round price, and by 0.32 percentage points ($p < 0.001$) when averaged across rounds. Effects on belief-price gaps are similar in magnitude, indicating that agents fail to move prices toward their beliefs in long-horizon markets, consistent with the opportunity cost mechanism discouraging informed trading.

These results replicate the result that time horizon reduces prediction market accuracy. However, the magnitude of the effect is markedly smaller than both theoretical and empirical benchmarks. The extended Antweiler model discussed in Subsection \ref{sec:hypotheses:subsec:H1} predicts a bias of approximately 6 percentage points at a two-year horizon with true probabilities of 0.05 or 0.95. The empirical estimates from \citet{page2013prediction} find biases of 4.7--10.9 percentage points for markets with more than 100 days to expiration. In contrast, the observed long-horizon effect on final-round price accuracy is only 0.72 percentage points, roughly one-tenth the magnitude predicted by theory.

The interaction terms test H2. For final-round price accuracy, the interaction coefficient is $-$0.60 percentage points ($p = 0.022$), indicating that interest payments reduce the horizon effect by approximately 83\% (from 0.72 to 0.12 percentage points). The residual horizon effect with interest ($\beta_1 + \beta_3 = 0.0012$) is economically negligible, and as Figure \ref{fig:summary_bars} illustrates, long-horizon markets with interest fall well within the confidence intervals of short-horizon markets.

\subsection{Prediction Market Exposure}

Table \ref{tab:ols_exposure} presents OLS regression results for prediction market exposure, defined as the market value of an agent's prediction market contracts as a share of total wealth.

\begin{table}[htbp]
\centering
\caption{OLS Regression Results: Prediction Market Exposure}
\label{tab:ols_exposure}
\begin{tabular}{lcc}
\toprule
& Last Round Exposure & Mean Exposure \\
\midrule
Intercept & 0.749*** & 0.590*** \\
& (0.013) & (0.008) \\
Long Horizon & $-$0.580*** & $-$0.369*** \\
& (0.018) & (0.012) \\
Interest & $-$0.005 & 0.001 \\
& (0.018) & (0.012) \\
Long $\times$ Interest & 0.451*** & 0.340*** \\
& (0.026) & (0.017) \\
\midrule
$R^2$ & 0.612 & 0.567 \\
$N$ & 400 & 400 \\
\bottomrule
\end{tabular}
\parbox{\textwidth}{\footnotesize\textit{Notes:} OLS regressions of prediction market exposure on treatment indicators. Standard errors in parentheses. * $p<0.10$, ** $p<0.05$, *** $p<0.01$.}
\end{table}

The effect on prediction market exposure is significant and economically substantial, strongly confirming H1. Agents in long-horizon markets without interest hold 58 percentage points less of their wealth in prediction market contracts compared to short-horizon markets ($p < 0.001$) in the last period, and 36.9 percentage points less on average. This large reduction in capital commitment directly reflects the opportunity cost mechanism: when outside options offer considerable returns over a two-year horizon, agents allocate less capital to prediction markets.

The interaction term of 45.1 percentage points ($p < 0.001$) indicates that interest payments offset 78\% of the horizon effect, supporting H2. With interest payments, agents in long-horizon markets hold 61.6\% of their wealth in prediction market contracts at the end of the final round, compared with 74.4\% in short-horizon markets. The residual horizon effect ($\beta_1 + \beta_3 = -0.129$) represents a 22\% reduction from the no-interest horizon penalty.

\subsection{Trading Volume}

Table \ref{tab:ols_volume} presents OLS regression results for total trading volume.

\begin{table}[htbp]
\centering
\caption{OLS Regression Results: Trading Volume}
\label{tab:ols_volume}
\begin{tabular}{lc}
\toprule
& Total Volume (in thousands of \$) \\
\midrule
Intercept & 135*** \\
& (3.37) \\
Long Horizon & $-$15.2*** \\
& (4.76) \\
Interest & 8.35* \\
& (4.76) \\
Long $\times$ Interest & 29.0*** \\
& (6.73) \\
\midrule
$R^2$ & 0.079 \\
$N$ & 400 \\
\bottomrule
\end{tabular}
\parbox{\textwidth}{\footnotesize\textit{Notes:} OLS regression of total trading volume on treatment indicators. Results displayed in thousands of dollars. Standard errors in parentheses. * $p<0.10$, ** $p<0.05$, *** $p<0.01$.}
\end{table}

Trading volume is significantly lower in long-horizon markets without interest, declining by \$15,182 ($p = 0.002$), consistent with H1. The interaction term of \$28,986 ($p < 0.001$) more than offsets this decline, supporting H2. Indeed, with interest payments, long-horizon markets exhibit higher volume than short-horizon markets ($\beta_1 + \beta_3 = \$13,804$), suggesting that interest payments successfully motivate active trading even over long time horizons.

\subsection{Bid-Ask Spreads}

Table \ref{tab:ols_spread} presents OLS regression results for bid-ask spreads, a measure of market liquidity.

\begin{table}[htbp]
\centering
\caption{OLS Regression Results: Bid-Ask Spread}
\label{tab:ols_spread}
\begin{tabular}{lcc}
\toprule
& Last Round Spread & Mean Spread \\
\midrule
Intercept & 0.0131*** & 0.0119*** \\
& (0.002) & (0.000) \\
Long Horizon & 0.0142*** & 0.0049*** \\
& (0.004) & (0.001) \\
Interest & $-$0.0017 & $-$0.0009 \\
& (0.003) & (0.001) \\
Long $\times$ Interest & $-$0.0027 & $-$0.0022** \\
& (0.005) & (0.001) \\
\midrule
$R^2$ & 0.050 & 0.140 \\
$N$ & 400 & 400 \\
\bottomrule
\end{tabular}
\parbox{\textwidth}{\footnotesize\textit{Notes:} OLS regressions of bid-ask spread on treatment indicators. Standard errors in parentheses. * $p<0.10$, ** $p<0.05$, *** $p<0.01$.}
\end{table}

Bid-ask spreads are significantly wider in long-horizon markets without interest, increasing by 1.42 percentage points in the final round ($p < 0.001$) and 0.49 percentage points on average ($p < 0.001$), consistent with H1. The interaction term for last round spread is $-$0.27 percentage points, indicating partial mitigation. However, it is not significant ($p = 0.585$). The interaction term for average spread is $-0.22$ percentage points and is significant at the 5\% level ($p = 0.017$). A residual horizon effect remains, even with interest, long-horizon markets exhibit spreads 1.15 percentage points wider than short-horizon markets ($\beta_1 + \beta_3 = 0.0115$). This suggests that while interest payments improve liquidity, they do not fully eliminate the horizon effect on market depth.

\subsection{Market Dynamics}

\begin{figure}[htbp]
\centering
\includegraphics[width=1.0\textwidth]{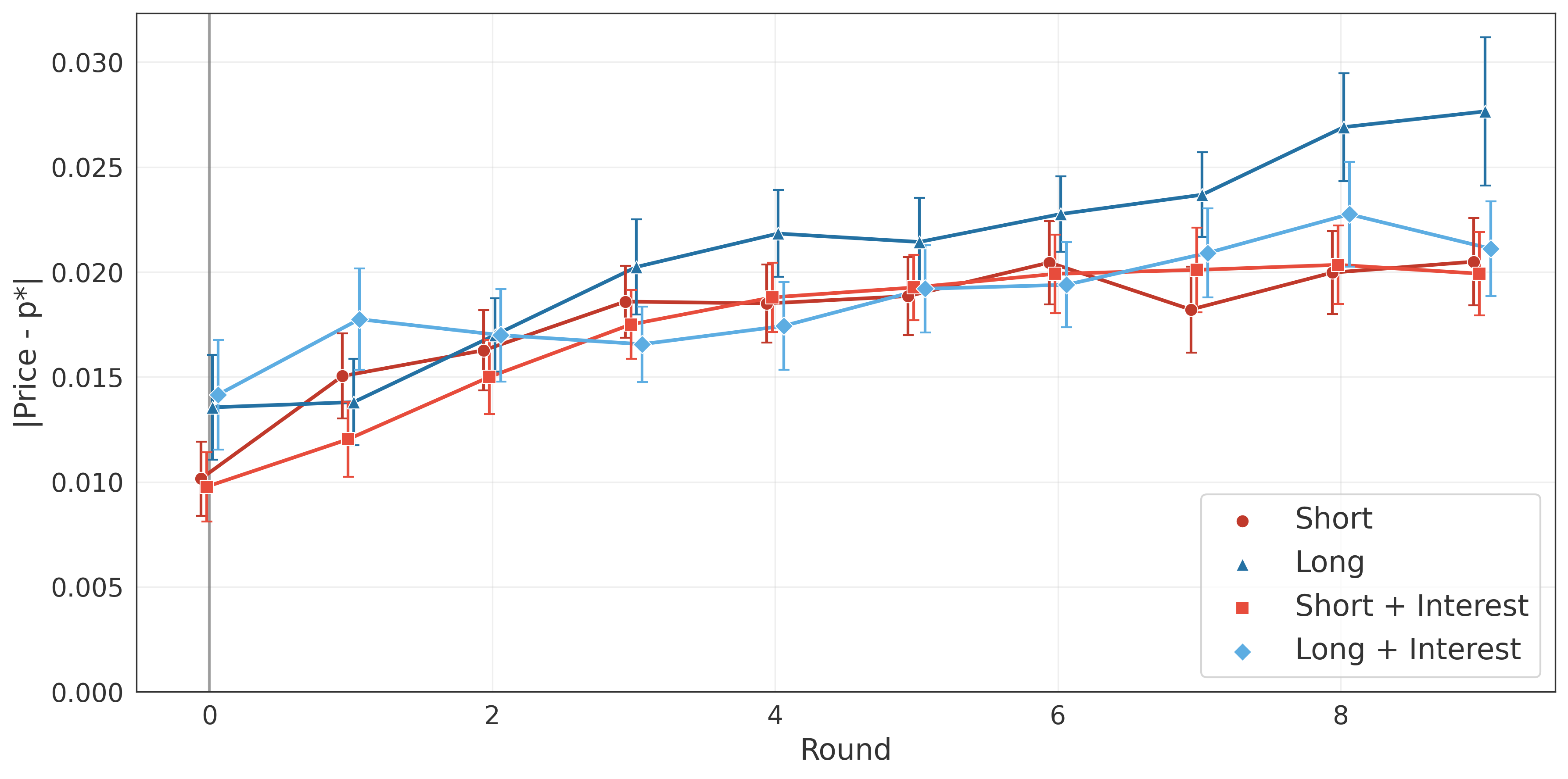}
\caption{Price Deviation from True Probability by Round}
\label{fig:price_deviation}
\floatfoot{\textit{Notes:} Absolute error between last trade price and true probability $p^*$, by trading round and treatment. Error bars show 95\% confidence intervals. $N = 100$ sessions per treatment.}
\end{figure}

The round-by-round evolution of market outcomes provides additional insight into how the treatments affect market behavior. Figure \ref{fig:price_deviation} displays the mean absolute error between market prices and true probabilities across trading rounds. Several patterns emerge. First, all treatments show increasing pricing errors over rounds, suggesting that early price discovery gives way to drift as trading continues. Second, the long-horizon treatment without interest consistently exhibits the highest pricing errors throughout the trading period, with the gap widening in later rounds. Third, the interest treatment brings long-horizon market performance close to that of short-horizon markets, particularly in the final rounds.

\begin{figure}[htbp]
\centering
\includegraphics[width=1.0\textwidth]{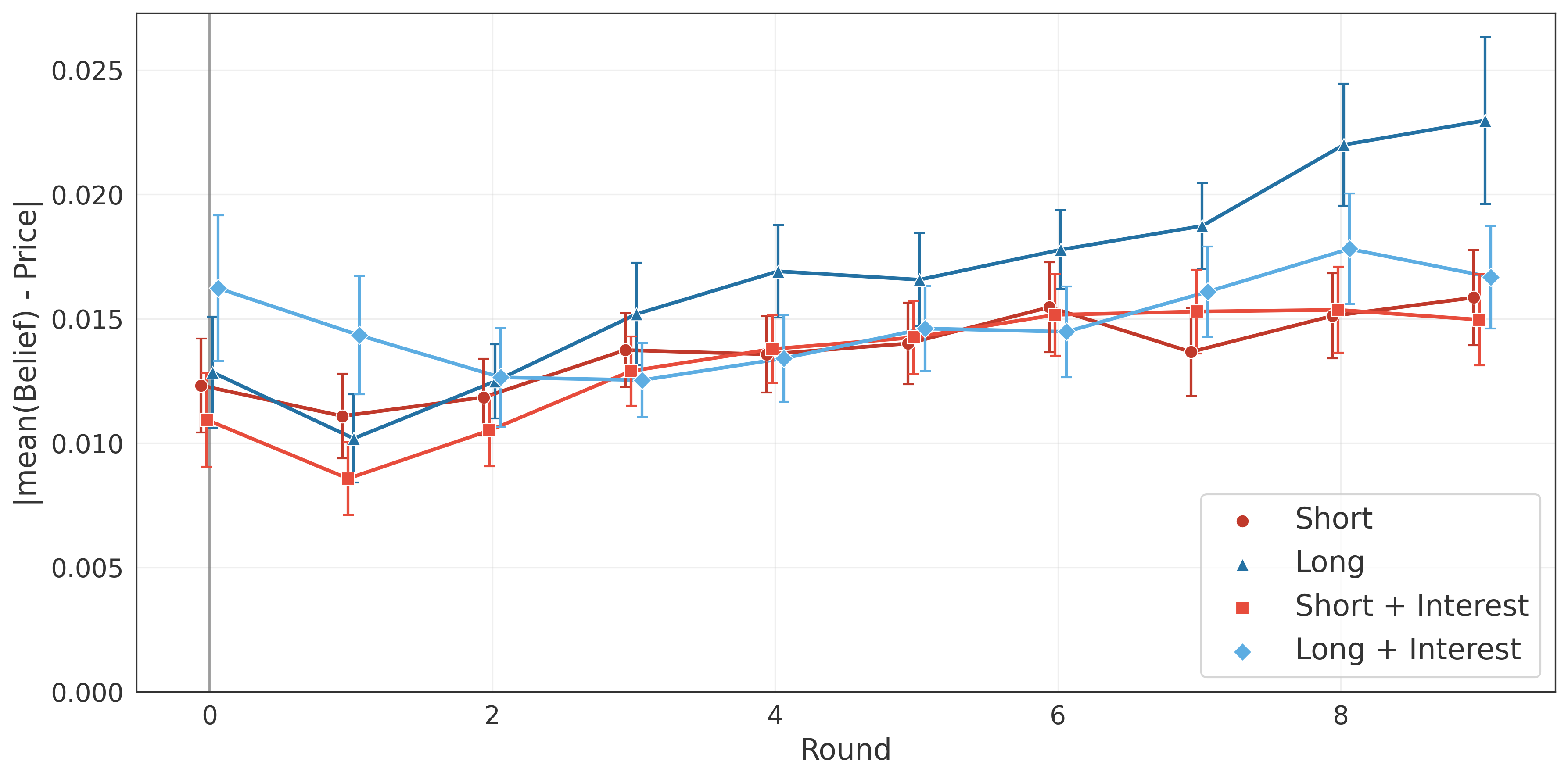}
\caption{Belief-Price Gap by Round}
\label{fig:belief_price_gap}
\floatfoot{\textit{Notes:} Absolute error between average agent belief and last trade price, by trading round and treatment. Error bars show 95\% confidence intervals. $N = 100$ sessions per treatment.}
\end{figure}

Figure \ref{fig:belief_price_gap} shows the gap between agents' average beliefs and market prices. The pattern is similar to the price-truth deviation: long-horizon markets without interest exhibit persistently larger belief-price gaps, and interest payments reduce this gap. The fact that belief-price gaps grow over time, particularly in the long-horizon no-interest treatment, suggests that agents increasingly choose not to trade on their information as rounds progress, consistent with agents recognizing that their capital could be earning returns elsewhere and becoming less willing to commit it to prediction market positions.

\begin{figure}[htbp]
\centering
\includegraphics[width=1.0\textwidth]{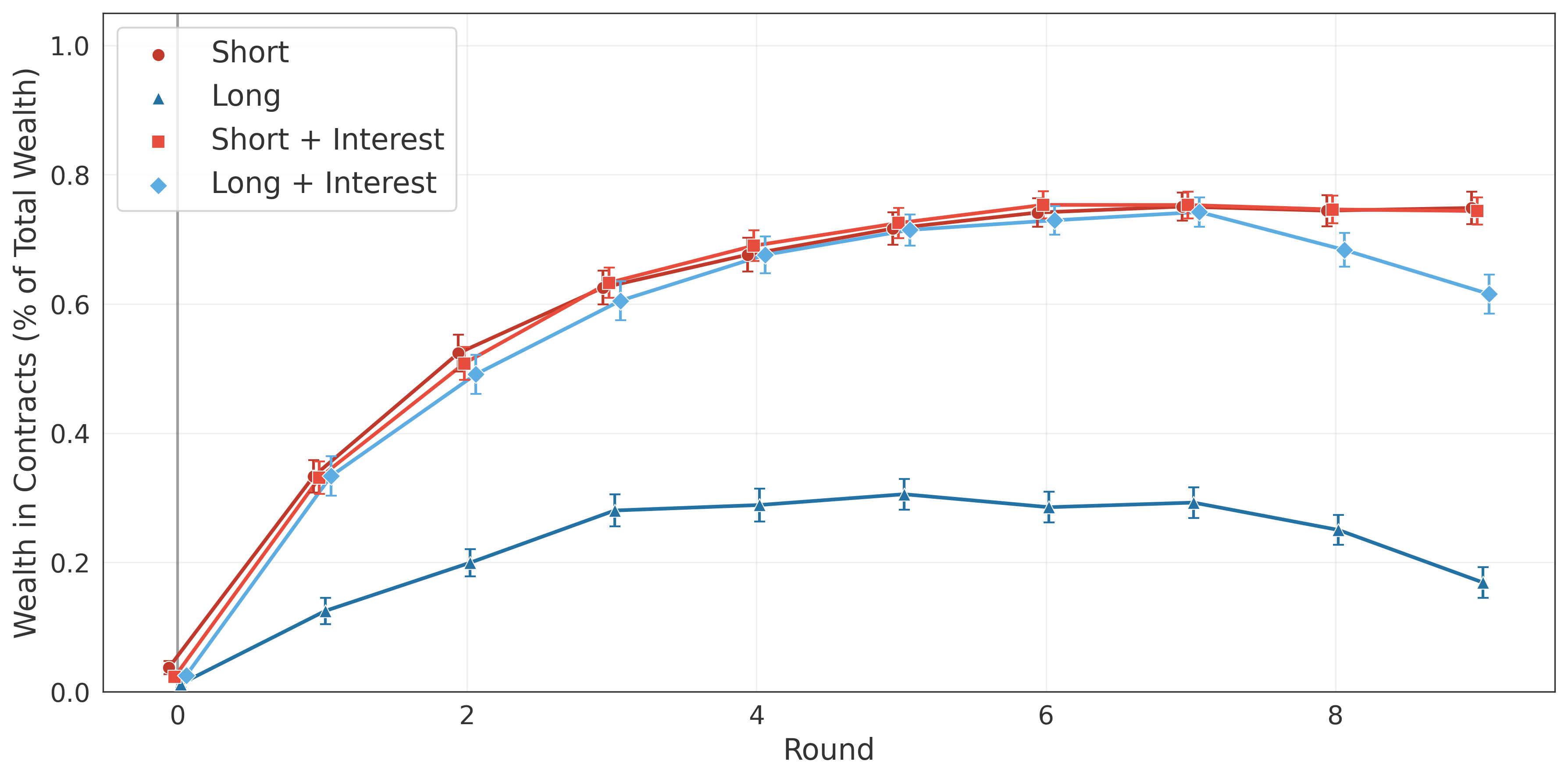}
\caption{Prediction Market Exposure by Round}
\label{fig:pm_wealth}
\floatfoot{\textit{Notes:} Mean fraction of total wealth held in prediction market contracts, by trading round and treatment. Error bars show 95\% confidence intervals. $N = 100$ sessions per treatment.}
\end{figure}

Figure \ref{fig:pm_wealth} displays prediction market exposure across rounds. The most salient feature is the large divergence between the long-horizon no-interest treatment and all other treatments. In short-horizon markets, agents rapidly increase their prediction market exposure in early rounds, reaching approximately 75\% of wealth by round 6. Long-horizon markets with interest follow a similar trajectory, though with a slight decline in the final rounds. In contrast, long-horizon markets without interest show agents maintaining exposure below 30\% throughout the trading period, with a notable decline in the final rounds as agents reduce positions before the post-trading allocation decision.

\begin{figure}[htbp]
\centering
\includegraphics[width=1.0\textwidth]{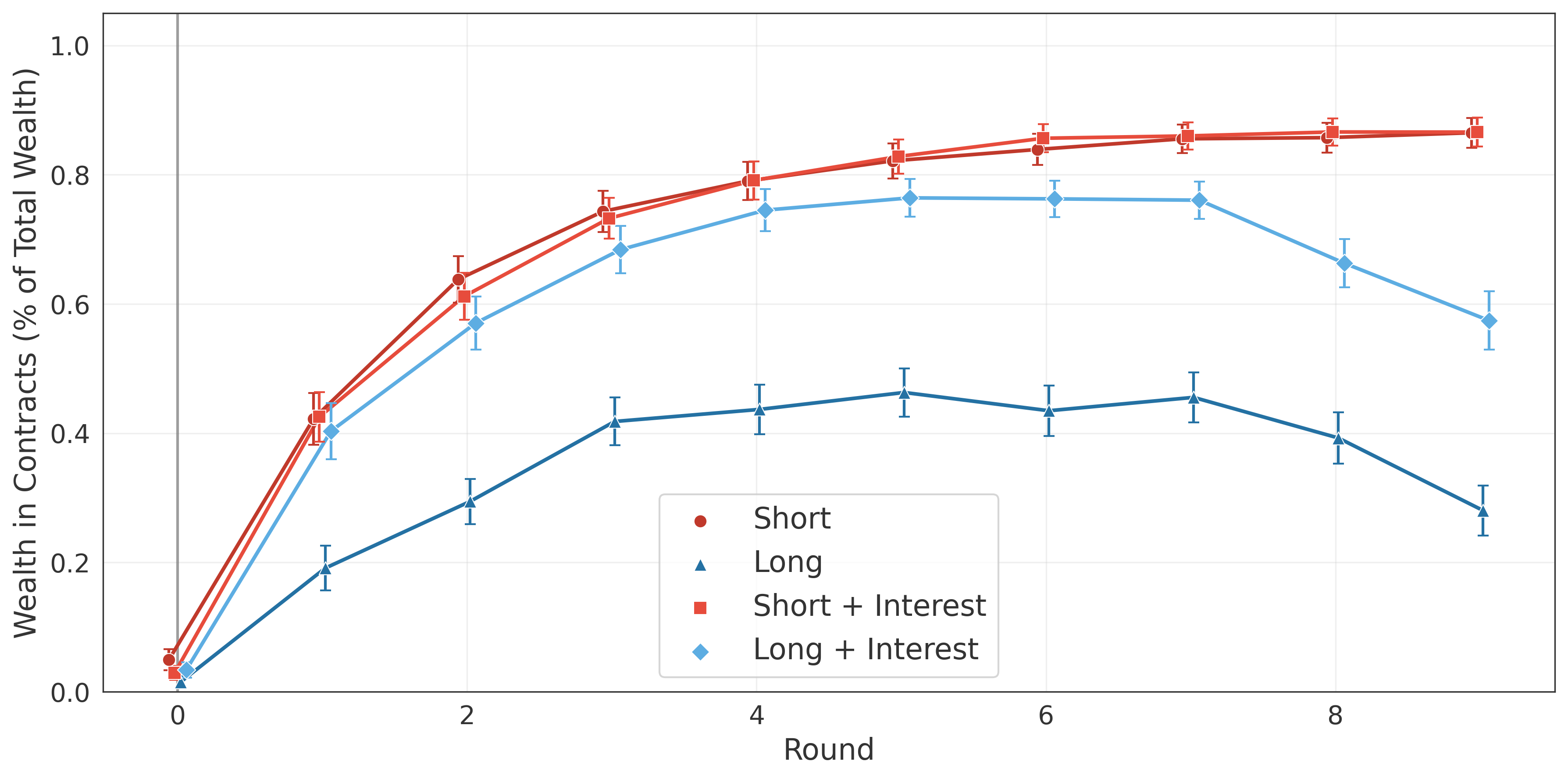}
\caption{Prediction Market Exposure by Round: High Risk Tolerance Agents}
\label{fig:pm_wealth_high}
\floatfoot{\textit{Notes:} Mean fraction of total wealth held in prediction market contracts for agents with high risk tolerance, by trading round and treatment. Error bars show 95\% confidence intervals.}
\end{figure}

\begin{figure}[htbp]
\centering
\includegraphics[width=1.0\textwidth]{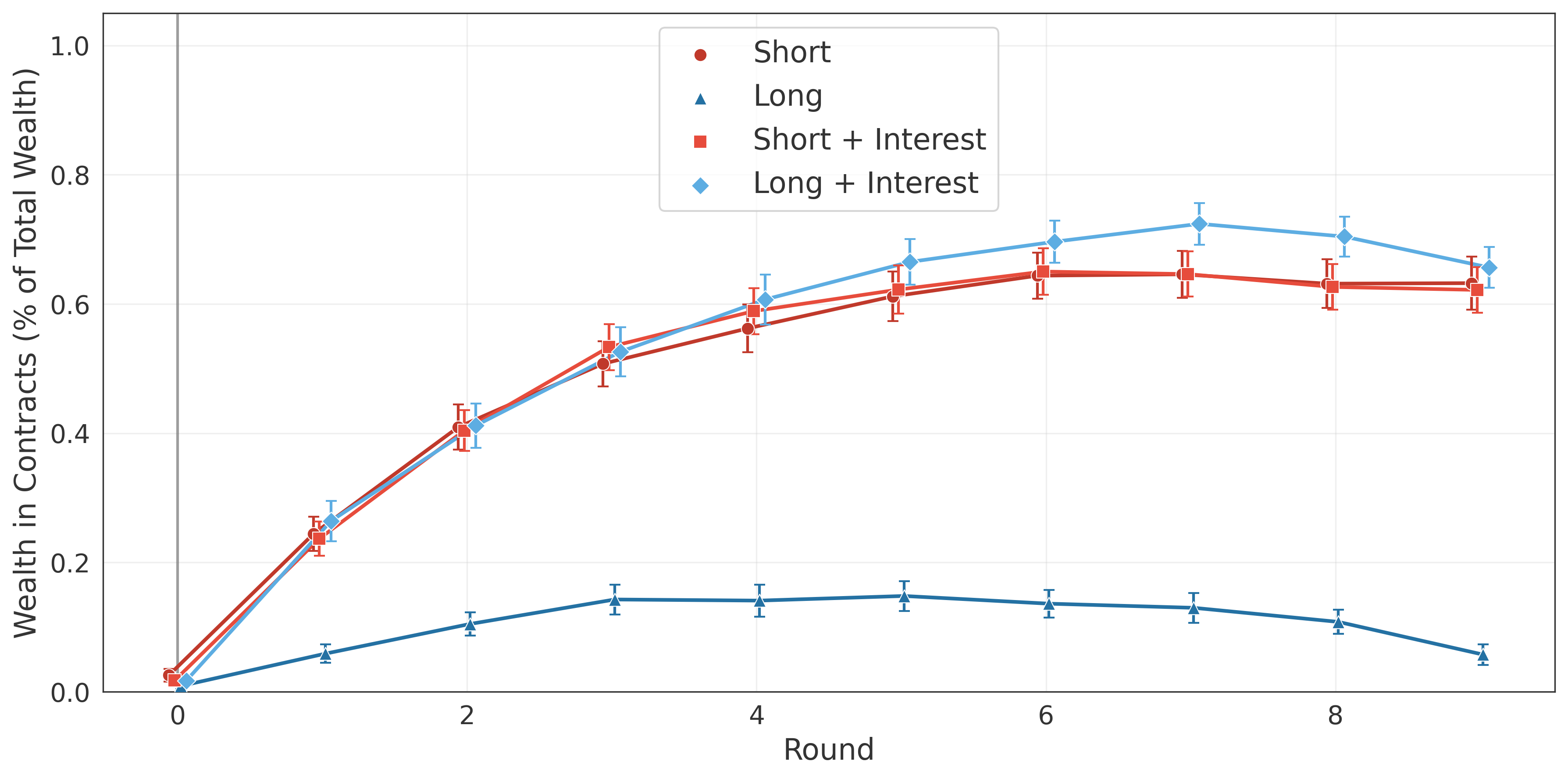}
\caption{Prediction Market Exposure by Round: Medium Risk Tolerance Agents}
\label{fig:pm_wealth_medium}
\floatfoot{\textit{Notes:} Mean fraction of total wealth held in prediction market contracts for agents with medium risk tolerance, by trading round and treatment. Error bars show 95\% confidence intervals.}
\end{figure}

This pattern is even more pronounced when examining agents by risk tolerance. Figure \ref{fig:pm_wealth_high} shows prediction market exposure for high risk tolerance agents, while Figure \ref{fig:pm_wealth_medium} shows the same for medium risk tolerance agents. High risk tolerance agents in all treatments except long-horizon no-interest reach exposure levels above 80\%, while medium risk tolerance agents plateau around 65\%. In the long-horizon no-interest treatment, however, both agent types maintain considerably lower exposure: high risk tolerance agents remain below 50\%, and medium risk tolerance agents below 20\%. This demonstrates that the opportunity cost mechanism affects all agent types, though its magnitude varies with risk preferences.

\begin{figure}[htbp]
\centering
\includegraphics[width=1.0\textwidth]{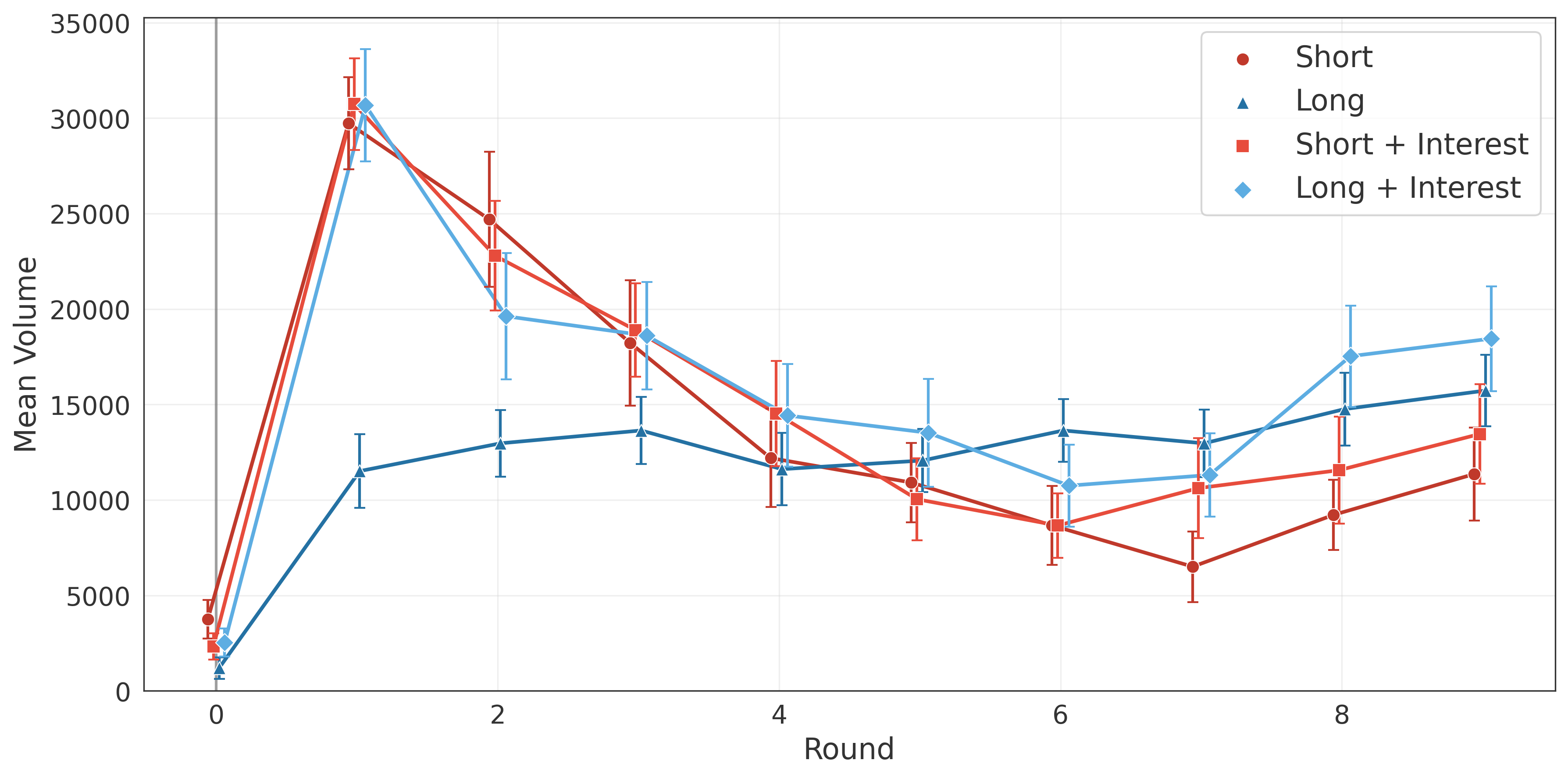}
\caption{Trading Volume by Round}
\label{fig:volume}
\floatfoot{\textit{Notes:} Mean trading volume in dollars, by trading round and treatment. Error bars show 95\% confidence intervals. $N = 100$ sessions per treatment.}
\end{figure}

Figure \ref{fig:volume} shows trading volume by round. Volume peaks in round 1 across all treatments as agents establish initial positions based on their private signals, then declines as the market approaches equilibrium. The long-horizon no-interest treatment exhibits notably lower volume throughout, consistent with reduced participation. Interestingly, the long-horizon with-interest treatment shows volume comparable to or slightly exceeding the short-horizon treatments, suggesting that interest payments successfully motivate active trading even over long time horizons.

\section{Conclusion} \label{sec:conclusion}

This paper provides the first experimental evaluation of interest-bearing positions as a solution to the long-horizon problem in prediction markets. Using LLM agent-based simulations in a controlled $2 \times 2$ factorial design, I find that both hypotheses are supported by the data: long horizons reduce market quality, and paying interest on positions mitigates this effect.

The most striking finding is the discrepancy between the observed magnitude of the long-horizon pricing bias and prior theoretical and empirical benchmarks. The extended Antweiler model predicts a bias of approximately 6 percentage points at a two-year horizon with extreme probabilities, and \citet{page2013prediction} find empirical biases of 4.7--10.9 percentage points for markets with more than 100 days to expiration. In contrast, the observed horizon effect on final-round price accuracy is only 0.72 percentage points, roughly one-tenth the theoretical prediction. This suggests that while opportunity costs do affect prediction market pricing, their impact on price accuracy may be considerably overstated in the existing literature.

Several factors may explain this discrepancy. First, my experimental design holds information quality constant across time horizons, eliminating the confound present in observational data where long-horizon markets often involve genuinely more uncertain events. Much of the apparent mispricing in long-horizon markets may reflect legitimate uncertainty rather than opportunity cost-induced bias. Second, LLM agents may respond differently to opportunity cost considerations than human traders, though evidence suggests LLMs if anything exhibit more rational behavior than humans in economic experiments \citep{chen2023emergence}, which should amplify rather than attenuate the opportunity cost mechanism.

While the effect of interest on pricing bias is economically modest, reducing an already small bias by approximately 83\%, the effect on market participation is substantial and may represent the more policy-relevant finding. In long-horizon markets without interest, agents allocated only 16.9\% of their wealth to prediction market positions, compared to 74.9\% in short-horizon markets. Introducing portfolio value interest increased long-horizon market exposure to 61.6\%. 

This dramatic increase in capital commitment has important implications. Recent work by \citet{rasooly2025manipulable} demonstrates that prediction market manipulations can remain effective for extended periods, with thinner markets being easier to manipulate. By increasing participation and trading volume, interest-bearing positions may make long-horizon markets more resilient to manipulation attempts. This protective effect could prove more valuable than the modest improvement in pricing accuracy, particularly for prediction markets addressing consequential long-term questions such as climate outcomes, technological developments, or geopolitical events where manipulation incentives may be strong.

These findings carry direct implications for prediction market platform design. The recent adoption of interest-bearing positions by Kalshi and Polymarket appears well-justified: even if the pricing accuracy benefits are smaller than theory suggests, the participation benefits are substantial. Platform operators considering whether to implement such policies can be reasonably confident that doing so will increase trader engagement with long-horizon markets. The interest rate need not exceed the risk-free rate to achieve these benefits, as the policy operates primarily by reducing the opportunity cost gap rather than eliminating it entirely.

Several limitations warrant discussion. First, while LLM agents have demonstrated human-like behavior across many economic domains, external validity remains an open question for market experiments. The finding that LLM-populated markets are ``too rational'' to generate bubbles \citep{henning2025llmagentsreplicatehuman} suggests that some behavioral phenomena may not translate. However, for the opportunity cost mechanism, which operates through rational capital allocation decisions rather than behavioral biases, LLM rationality may provide a cleaner test of the theoretical mechanism. Second, the experimental design examines a single binary prediction market rather than the portfolio of markets traders face in practice, where opportunity cost considerations may interact across contracts. Third, the two-year horizon, while long by prediction market standards, remains short relative to questions about climate change or other multi-decade events where the long-horizon problem may be most severe.

Future research could extend this analysis in several directions. Human subject experiments could validate whether the findings generalize beyond LLM agents, though the cost and time requirements of long-horizon market experiments with human participants are substantial. Field experiments manipulating interest policies on actual prediction market platforms could provide direct evidence on real-world effectiveness, though such experiments face obvious practical and ethical constraints. Theoretical work could explore whether modified interest structures, such as interest rates that increase with time to resolution or interest paid on committed capital rather than mark-to-market portfolio value, might further improve long-horizon market performance.

In conclusion, this paper contributes to the prediction market literature by demonstrating that interest-bearing positions effectively address the long-horizon problem, primarily through their impact on market participation rather than pricing accuracy. The finding that opportunity cost-induced pricing bias is smaller than previously believed provides a more optimistic outlook for long-horizon prediction markets. While they will always face participation challenges, these challenges may not compromise their core function of aggregating information into accurate probability forecasts.

\setlength\bibsep{0pt}
\bibliographystyle{apalike}
\bibliography{references}

\clearpage

\appendix
\section*{Appendix}

\section{Extension of Antweiler's Opportunity Cost Model} \label{sec:appendixa}

This appendix extends the theoretical framework of \citet{antweiler2012long} to incorporate the specific features of my experimental design: heterogeneous risk preferences among traders and the availability of a risky outside investment option alongside the risk-free asset.

\subsection{Antweiler's Original Framework}

\citet{antweiler2012long} models traders with exponential utility who face an opportunity cost when committing capital to prediction markets. In his framework, a trader $i$ with belief $\hat{p}_i$ about the probability of the event considers whether to take a position at market price $p$. The key insight is the asymmetry between long and short positions: a long position (buying YES contracts) at price $p$ puts $p$ dollars at risk per contract, while a short position (equivalently, buying NO contracts) puts $(1-p)$ dollars at risk.

In the presence of an outside investment opportunity earning return $r$ over horizon $T$, the opportunity cost creates a participation constraint. Defining $\theta = e^{rT} - 1$ as the cumulative return and $\psi = \theta/(1+\theta)$ as the transformed opportunity cost parameter, a trader will only take a long position if $1 - p > \psi$, and will only take a short position if $p > \psi$. This creates a ``no-trade region'' around the market price where traders with moderate beliefs abstain.

Building on the log-odds demand framework of \citet{wolfers2006interpreting}, Antweiler derives the demand for contracts by trader $i$ adjusted for opportunity costs:
\begin{equation}
q_i = \frac{w_i}{\alpha_i} \left[ \ln\left(\frac{\hat{p}_i}{1-\hat{p}_i}\right) - \ln\left(\frac{p - \psi \cdot \mathbf{1}(q_i < 0)}{1 - p - \psi \cdot \mathbf{1}(q_i > 0)}\right) \right]
\end{equation}
where $w_i$ is initial wealth and $\mathbf{1}(\cdot)$ is the indicator function. The equilibrium price is determined by market clearing: $\sum_i q_i = 0$.

\subsection{Extension to Heterogeneous Risk Preferences and Risky Outside Options}

My experimental design introduces two features absent from Antweiler's baseline model. First, traders have heterogeneous risk preferences rather than a common discount rate. Second, traders can allocate their outside wealth between a risk-free asset (earning $r_f$) and a risky asset (with expected return $\mu_m$ and volatility $\sigma_m$).

I model trader heterogeneity through a distribution of risk aversion parameters $\alpha_i \sim \text{Lognormal}(\mu_\alpha, \sigma_\alpha)$. Each trader optimally allocates between the risk-free and risky assets according to the standard mean-variance portfolio choice:
\begin{equation}
\omega_i^* = \frac{\mu_m - r_f}{\alpha_i \sigma_m^2}
\end{equation}
where $\omega_i^*$ is the fraction invested in the risky asset, constrained to $[0, 1]$ for no leverage or short-selling.

The trader's optimal outside return is therefore:
\begin{equation}
r_i = r_f + \omega_i^* (\mu_m - r_f)
\end{equation}

This generates trader-specific opportunity cost parameters:
\begin{equation}
\theta_i = e^{r_i T} - 1, \qquad \psi_i = \frac{\theta_i}{1 + \theta_i}
\end{equation}

Traders with low risk aversion invest more heavily in the risky asset, earn higher expected returns, and therefore face higher opportunity costs when committing capital to prediction markets.

\subsection{Belief Distribution}

In the experiment, each trader observes $n = 200$ independent Bernoulli draws with true probability $p^*$. Applying Bayesian updating with a $\text{Beta}(1,1)$ (uniform) prior, a trader observing $k$ successes forms posterior belief:
\begin{equation}
\hat{p}_i = \frac{1 + k}{2 + n}
\end{equation}

The number of successes follows a Binomial distribution: $k \sim \text{Binomial}(n, p^*)$. This generates a distribution of beliefs across traders that is centered near the true probability but exhibits dispersion due to sampling variation.

\subsection{Equilibrium Computation}

The equilibrium price is computed by integrating demand over the joint distribution of beliefs and risk aversion parameters. For a given price $p$, total demand is:
\begin{equation}
D(p) = \int_\alpha \int_k q(p, \hat{p}(k), \alpha, \psi(\alpha)) \cdot f(k | p^*) \cdot g(\alpha) \, dk \, d\alpha
\end{equation}
where $f(k | p^*)$ is the Binomial probability mass function and $g(\alpha)$ is the Lognormal density. The equilibrium price $p^*$ satisfies $D(p^*) = 0$.

I solve this numerically using a dense grid over the risk aversion distribution and exact Binomial probabilities for the belief distribution. The equilibrium is found via root-finding (Brent's method) on the interval $(0.001, 0.999)$.

\subsection{Quantitative Predictions}

Using the experimental parameters ($T = 2$ years, $r_f = 0.04$, $\mu_m = 0.10$, $\sigma_m = 0.16$, $n = 200$ signals, initial wealth $w = \$10{,}000$) and assuming risk aversion parameters are distributed $\text{Lognormal}(0.5, 0.6)$, the model generates the following predictions for a true probability of $p^* = 0.05$:

\begin{center}
\begin{tabular}{lc}
\toprule
Quantity & Value \\
\midrule
True event probability & 0.050 \\
Expected posterior belief & 0.05446 \\
Equilibrium market price & 0.11488 \\
Long-horizon bias ($p - \mathbb{E}[\hat{p}]$) & 0.06042 \\
\bottomrule
\end{tabular}
\end{center}

The predicted bias of approximately 6 percentage points toward 0.50 reflects the asymmetric participation constraints: at a true probability of 0.05, the short side of the market requires more capital at risk, deterring participation by traders who would otherwise push the price down toward the true value. Due to the symmetry of the model, the long-horizon bias at a true probability of 0.95 is symmetric and identical in magnitude.

\section{Experiment Transcript} \label{sec:appendixb}


\begin{tcolorbox}[title={System Prompt}, breakable, colback=gray!5, colframe=gray!75]
\textbf{\Large Prediction Market Experiment}\\

\par\medskip
You are participating in an economic experiment about prediction markets.

\par\medskip
\textbf{\large The Experiment Structure}\\

\par\medskip
There are 10 participants (including you). The experiment consists of \textbf{10 trading rounds}.

\par\medskip
At the end of trading, a binary event will either occur or not occur. The event will resolve in 2 years. The event has a true probability of occurring that is fixed and lies in the interval [0.0, 1.0]. This true probability is unknown to all participants.

\par\medskip
All participants start with \$10,000.00 in cash and no initial positions.

\par\medskip
\textbf{Private Signals}\\
At the start of the experiment, you (and the other participants) observe \textbf{200 independent draws} from a Bernoulli distribution with the TRUE event probability.
You will see how many of these draws were "successes".

\par\medskip
\textbf{Contracts}\\
You trade two types of contracts:
\begin{itemize}
  \item \textbf{YES contracts}: Pay \$1.00 if the event occurs, \$0.00 if it does not
  \item \textbf{NO contracts}: Pay \$1.00 if the event does NOT occur, \$0.00 if it does
\end{itemize}

\par\medskip
Contract prices range from \$0.00 to \$1.00.

\par\medskip
\textbf{Outside Options}\\
After trading concludes (but before the event resolves), you will have the option to invest any cash you have remaining in two assets:
\begin{itemize}
  \item \textbf{Risk-free asset}: Earns 8\% (guaranteed)
  \item \textbf{Risky asset}: Expected 20\% return, with a standard deviation of 22.63\%
\end{itemize}

\par\medskip
\textbf{Position Interest}\\
After trading concludes, you earn \textbf{8\% interest} on the market value of your prediction market contracts.
\begin{itemize}
  \item Interest = (YES contracts × YES price at end of last round + NO contracts × NO price at end of last round) × 8\%
  \item This interest is paid in cash after trading concludes.
\end{itemize}

\par\medskip
\textbf{Experiment Flow}\\
The experiment follows this sequence:
1. \textbf{You receive 200 private signals} about the true event probability at the start
2. \textbf{10 Trading Rounds}: You can place orders to buy and sell prediction market contracts
3. \textbf{Post-Trading Investment Decision}: After trading ends, you allocate your remaining cash between the risk-free asset and the risky asset
4. \textbf{Market Resolution}: Returns on outside investments and position interest are paid out. The event resolves and contracts pay out.

\par\medskip
\textbf{Trading}\\
Each round, you may place limit orders to buy or sell YES and NO contracts:
\begin{itemize}
  \item Orders execute immediately if matched, otherwise wait in the order book
  \item You can only sell contracts you own
  \item Orders persist until filled or cancelled
  \item Open buy orders reduce your cash available for trading. You may need to replace or cancel some orders to make room for new ones.
  \item Similarly, open sell orders reduce your contracts available to sell. You may need to replace or cancel some orders to make room for new ones.
  \item Order submission is shuffled each round before matching so no one has a fixed priority
\end{itemize}

\par\medskip
\textbf{\large Your Goal}\\

\par\medskip
Your objective is to \textbf{maximize your final wealth}.

\par\medskip
\textbf{\large Your Risk Profile}\\

\par\medskip
You have MEDIUM risk tolerance. You balance risk and return considerations, seeking reasonable returns while avoiding excessive risk. You make decisions that balance potential gains with downside protection.

\par\medskip
\textbf{\large Response Formats}\\

\par\medskip
There are two types of decisions you will make:

\par\medskip
\textbf{Trade Decisions (during trading rounds)}\\

\par\medskip
Each trading round, provide:

\par\medskip
\textbf{Required Fields:}
\begin{itemize}
  \item \textbf{probability\_estimate} (float 0-1): Your estimate of the event probability
  \item \textbf{orders} (list): Orders you want to place
  \item \textbf{replace\_decision} (string): How to handle existing orders
  \item "Add": Keep existing orders, add new ones (requires AVAILABLE resources)
  \item "Cancel": Cancel ALL existing orders
  \item "Replace": Your orders list becomes complete set of active orders
\end{itemize}

\par\medskip
\textbf{Order Format:}
\begin{itemize}
  \item \textbf{contract\_type}: "YES" or "NO"
  \item \textbf{decision}: "Buy" or "Sell"
  \item \textbf{quantity}: Number of contracts (integer)
  \item \textbf{price\_limit}: Limit price (float 0-1)
\end{itemize}

\par\medskip
\textbf{Post-Trading Investment Decision (after trading ends)}\\

\par\medskip
After trading is complete, you will decide how to allocate your remaining cash:

\par\medskip
\textbf{Required Fields:}
\begin{itemize}
  \item \textbf{risky\_allocation\_pct} (float 0-1): Percentage of your remaining cash to invest in the risky asset
  \item The remainder (1 - risky\_allocation\_pct) goes into the risk-free asset
\end{itemize}
\end{tcolorbox}

\begin{tcolorbox}[title={Round 1 Trading Decision - Prompt}, breakable, colback=blue!5, colframe=blue!75]
\textbf{\Large Round 1 trading round}\\

\par\medskip
\textbf{\large Timing}\\
\begin{itemize}
  \item \textbf{Round}: 1 of 10
\end{itemize}
\textbf{\large Your Private Signals}\\
You observed \textbf{12 successes} out of \textbf{200 draws}.

\par\medskip
\textbf{\large Current Market State}\\

\par\medskip
\textbf{YES Contract Market}\\

\par\medskip
\textbf{\small Best Prices}\\
\begin{itemize}
  \item Best Bid (sell): None
  \item Best Ask (buy): None
\end{itemize}

\par\medskip
\textbf{\small Order Book}\\
\textbf{\footnotesize Sell Orders (Asks)}\\
\textit{No sell orders}
\textbf{\footnotesize Buy Orders (Bids)}\\
\textit{No buy orders}

\par\medskip
\textbf{NO Contract Market}\\

\par\medskip
\textbf{\small Best Prices}\\
\begin{itemize}
  \item Best Bid (sell): None
  \item Best Ask (buy): None
\end{itemize}

\par\medskip
The No order book is the inverse of the Yes book: a YES buy order at \$0.23 is equivalent to a NO sell order at \$0.77.

\par\medskip
\textbf{Your Current Position}\\

\par\medskip
\begin{itemize}
  \item \textbf{Total Cash}: \$10000.00
  \item \textbf{AVAILABLE for new orders: \$10000.00}
  \item \textit{Cash remaining after trading can be invested in a risk-free asset (8\% return) or a risky asset (expected 20\% return, 22.63\% std dev)}
  \item \textbf{YES Contracts}: 0 total (current value: \$0.00)
  \item \textbf{AVAILABLE for new sell orders: 0}
  \item \textbf{NO Contracts}: 0 total (current value: \$0.00)
  \item \textbf{AVAILABLE for new sell orders: 0}
\end{itemize}

\par\medskip
\textit{Reminder: After trading concludes, you will earn 8\% interest on the market value of your contract holdings (based on end-of-trading prices).}

\par\medskip
\textbf{\small Your Outstanding Orders}\\
\textit{No outstanding orders}

\par\medskip
What is your trading decision?
\end{tcolorbox}

\begin{tcolorbox}[title={Round 1 Trading Decision - Response}, breakable, colback=green!5, colframe=green!75]
\textbf{Reasoning Summary:}
\begin{quote}
\textbf{Evaluating probability estimates}

I'm figuring out the probability estimates and order placements. First, I've calculated a private signal of 12 out of 200, which provides a probability of about 0.06. With a uniform prior and posterior of Beta(13,189), the mean works out to around 0.064356. This allows for the possibility of placing bids and asks around this estimate, noticing we can't sell unless we own the asset. We could consider buying NO at 0.90, giving us exposure to a high probability of it not occurring.
\end{quote}
\begin{quote}
\textbf{Analyzing expected value and bids}

So, buying at 0.90 gives an expected value of 0.036 per contract if the probability is 0.936. However, if we overpay, it’s better to place bids lower to capture any mispricing. I can anchor around 0.06 for the YES price, making the equivalent NO price 0.94. I think we should bid YES at around 0.04-0.05 and also consider bidding NO at 0.92, which reflects a fair assessment of YES being overpriced at 0.08. Buying NO at 0.92 seems like a good undervalued opportunity!
\end{quote}
\begin{quote}
\textbf{Assessing fair value and risk}

I'm realizing that the fair value for NO is 0.936, so paying 0.92 is beneficial. This means I can place bids for NO around 0.92 or 0.93, but there’s a chance that buyers might scoop them up at 0.07-0.08. Buying YES at 0.05 seems reasonable for a small amount, but I don’t want a long position unless it's really cheap. With a starting bankroll of 10k, each NO contract could yield an ROI of 8.7\%, but there's a significant risk of loss if the event occurs, so I shouldn't bet all at once.
\end{quote}
\begin{quote}
\textbf{Planning allocation and trades}

I'm thinking about allocating around 30\% to this bet. If I buy 3,000 NO contracts at 0.92, it would cost 2,760, and the loss would be manageable if the event occurs. The expected gain seems small, about 48. If I go for 5,000 contracts at 0.92, the cost would be 4,600 with an expected gain of 80. I can set limit orders at better prices like 0.85 to catch any opportunities without risking it all. 

I plan to place an order for:

- Buy 3,000 NO at 0.90, still below the fair price of 0.936.
- Buy 2,000 NO at 0.88 for a better deal.
- Buy 500 YES at 0.04 as a low-cost option. 

I’ll ensure the total reserve is acceptable for cash flow. If filled, I'd have a significant NO position, which seems fine. I won’t set sell orders since I don't own any yet.
\end{quote}
\begin{quote}
\textbf{Updating probability estimate}

I have a probability estimate of 0.064. It’s clear I need to replace a decision, since there's no existing one at the moment. I'll also provide an updated orders list to reflect this change. This new information will shape my future actions effectively. Let’s get everything aligned with this estimate and ensure all my orders are correctly set up to optimize the trading strategy.
\end{quote}

\textbf{Response:}
\begin{verbatim}
{
  "probability_estimate": 0.0644,
  "orders": [
    {
      "decision": "Buy",
      "quantity": 3000,
      "contract_type": "NO",
      "order_type": "limit",
      "price_limit": 0.9
    },
    {
      "decision": "Buy",
      "quantity": 2000,
      "contract_type": "NO",
      "order_type": "limit",
      "price_limit": 0.88
    },
    {
      "decision": "Buy",
      "quantity": 500,
      "contract_type": "YES",
      "order_type": "limit",
      "price_limit": 0.04
    }
  ],
  "replace_decision": "Add"
}
\end{verbatim}
\end{tcolorbox}

\begin{tcolorbox}[title={Post-Trading Investment Decision - Prompt}, breakable, colback=blue!5, colframe=blue!75]
\textbf{\Large Post-Trading Investment Decision}\\

\par\medskip
Trading has ended. Now you must decide how to invest your remaining cash.

\par\medskip
\textbf{\large Your Investment Options}\\
\begin{itemize}
  \item \textbf{Risk-free asset}: Earns 8\% (guaranteed)
  \item \textbf{Risky asset}: Expected 20\% return, with a standard deviation of 22.63\%
\end{itemize}

\par\medskip
\textbf{\large Your Current Financial State}\\
\begin{itemize}
  \item \textbf{Remaining Cash}: \$6295.65
  \item \textbf{YES Contracts Held}: 56257.00 (current value: \$3375.42) (Earns 8\% on market value after trading)
  \item \textbf{NO Contracts Held}: 0.00 (current value: \$0.00) (Earns 8\% on market value after trading)
\end{itemize}

\par\medskip
Note: Your contracts will pay out when the event resolves. This decision is only about how to invest your remaining cash.

\par\medskip
What percentage of your remaining cash do you want to invest in the risky asset?
\end{tcolorbox}

\begin{tcolorbox}[title={Post-Trading Investment Decision - Response}, breakable, colback=green!5, colframe=green!75]
\textbf{Response:}
\begin{verbatim}
{
  "risky_allocation_pct": 0.55
}
\end{verbatim}
\end{tcolorbox}

\section{Robustness Checks} \label{sec:appendixc}

This appendix reports robustness checks for the main results. Section C.1 presents non-parametric Mann-Whitney U tests corresponding to the OLS regressions in the main text. Section C.2 reports results using the bid-ask midpoint as the price measure rather than the last trade price.

\subsection{Non-Parametric Tests}

While the main text reports OLS regressions with interpretable coefficient estimates, the experimental economics literature often employs non-parametric tests that make fewer distributional assumptions. Tables \ref{tab:mwu_ltp_h1}--\ref{tab:mwu_ltp_h2_residual} report one-sided Mann-Whitney U tests for all primary outcomes using the last trade price. The results are consistent with the OLS findings.

\begin{table}[htbp]
\centering
\caption{Mann-Whitney U Tests: H1 Long-Horizon Effect (Cell 2 vs. Cell 1)}
\label{tab:mwu_ltp_h1}
\begin{tabular}{llrr}
\toprule
Outcome & Predicted Direction & $U$ Statistic & $p$-value \\
\midrule
Price-Truth MAE (Last) & Cell 2 $>$ Cell 1 & 6188.0 & 0.002 \\
Price-Truth MAE (Mean) & Cell 2 $>$ Cell 1 & 6330.0 & $<$0.001 \\
Price-Belief MAE (Last) & Cell 2 $>$ Cell 1 & 6273.0 & $<$0.001 \\
Price-Belief MAE (Mean) & Cell 2 $>$ Cell 1 & 6588.0 & $<$0.001 \\
PM Exposure (Last) & Cell 2 $<$ Cell 1 & 19.0 & $<$0.001 \\
PM Exposure (Mean) & Cell 2 $<$ Cell 1 & 2.0 & $<$0.001 \\
Trading Volume & Cell 2 $<$ Cell 1 & 3638.0 & $<$0.001 \\
Bid-Ask Spread (Last) & Cell 2 $>$ Cell 1 & 4354.5 & $<$0.001 \\
Bid-Ask Spread (Mean) & Cell 2 $>$ Cell 1 & 7818.0 & $<$0.001 \\
\bottomrule
\end{tabular}
\parbox{\textwidth}{\footnotesize\textit{Notes:} One-sided Mann-Whitney U tests comparing Cell 2 (long horizon, no interest) to Cell 1 (short horizon, no interest). Each cell contains 100 independent simulation sessions.}
\end{table}

\begin{table}[htbp]
\centering
\caption{Mann-Whitney U Tests: H2 Direct Effect of Interest (Cell 4 vs. Cell 2)}
\label{tab:mwu_ltp_h2_direct}
\begin{tabular}{llrr}
\toprule
Outcome & Predicted Direction & $U$ Statistic & $p$-value \\
\midrule
Price-Truth MAE (Last) & Cell 4 $<$ Cell 2 & 3989.0 & 0.006 \\
Price-Truth MAE (Mean) & Cell 4 $<$ Cell 2 & 4008.5 & 0.008 \\
Price-Belief MAE (Last) & Cell 4 $<$ Cell 2 & 4027.0 & 0.009 \\
Price-Belief MAE (Mean) & Cell 4 $<$ Cell 2 & 4053.0 & 0.010 \\
PM Exposure (Last) & Cell 4 $>$ Cell 2 & 9863.0 & $<$0.001 \\
PM Exposure (Mean) & Cell 4 $>$ Cell 2 & 9988.0 & $<$0.001 \\
Trading Volume & Cell 4 $>$ Cell 2 & 7806.0 & $<$0.001 \\
Bid-Ask Spread (Last) & Cell 4 $<$ Cell 2 & 1830.0 & 0.010 \\
Bid-Ask Spread (Mean) & Cell 4 $<$ Cell 2 & 3316.0 & $<$0.001 \\
\bottomrule
\end{tabular}
\parbox{\textwidth}{\footnotesize\textit{Notes:} One-sided Mann-Whitney U tests comparing Cell 4 (long horizon, with interest) to Cell 2 (long horizon, no interest). Each cell contains 100 independent simulation sessions.}
\end{table}

\begin{table}[htbp]
\centering
\caption{Mann-Whitney U Tests: H2 Residual Horizon Effect (Cell 4 vs. Cell 3)}
\label{tab:mwu_ltp_h2_residual}
\begin{tabular}{llrr}
\toprule
Outcome & Predicted Direction & $U$ Statistic & $p$-value \\
\midrule
Price-Truth MAE (Last) & Cell 4 $>$ Cell 3 & 5175.5 & 0.333 \\
Price-Truth MAE (Mean) & Cell 4 $>$ Cell 3 & 5548.5 & 0.090 \\
Price-Belief MAE (Last) & Cell 4 $>$ Cell 3 & 5419.0 & 0.153 \\
Price-Belief MAE (Mean) & Cell 4 $>$ Cell 3 & 5846.0 & 0.019 \\
PM Exposure (Last) & Cell 4 $<$ Cell 3 & 2585.0 & $<$0.001 \\
PM Exposure (Mean) & Cell 4 $<$ Cell 3 & 4050.0 & 0.010 \\
Trading Volume & Cell 4 $<$ Cell 3 & 6235.0 & 0.999 \\
Bid-Ask Spread (Last) & Cell 4 $>$ Cell 3 & 4092.0 & $<$0.001 \\
Bid-Ask Spread (Mean) & Cell 4 $>$ Cell 3 & 7602.5 & $<$0.001 \\
\bottomrule
\end{tabular}
\parbox{\textwidth}{\footnotesize\textit{Notes:} One-sided Mann-Whitney U tests comparing Cell 4 (long horizon, with interest) to Cell 3 (short horizon, with interest). Each cell contains 100 independent simulation sessions.}
\end{table}

\subsection{Alternative Price Measure: Bid-Ask Midpoint}

The main text uses the last trade price as the primary price measure, following the convention employed by Kalshi. As a robustness check, I re-estimate all price accuracy results using the bid-ask midpoint, following the convention employed by Polymarket.

\subsubsection{OLS Results with Midpoint Price}

Table \ref{tab:ols_midpoint} presents OLS regression results for price accuracy measures using the bid-ask midpoint.

\begin{table}[htbp]
\centering
\caption{OLS Regression Results: Price Accuracy (Midpoint Price)}
\label{tab:ols_midpoint}
\begin{tabular*}{\textwidth}{@{\extracolsep{\fill}}lcccc}
\toprule
& \multicolumn{2}{c}{\textbf{Price-Truth MAE}} & \multicolumn{2}{c}{\textbf{Price-Belief MAE}} \\
\cmidrule(lr){2-3} \cmidrule(lr){4-5}
& Last Round & Mean & Last Round & Mean \\
\midrule
Intercept & 0.0207*** & 0.0195*** & 0.0156*** & 0.0150*** \\
& (0.002) & (0.001) & (0.002) & (0.001) \\
Long Horizon & 0.0040 & 0.0025** & 0.0038 & 0.0024*** \\
& (0.003) & (0.001) & (0.002) & (0.001) \\
Interest & $-$0.0013 & $-$0.0006 & $-$0.0007 & $-$0.0006 \\
& (0.002) & (0.001) & (0.002) & (0.001) \\
Long $\times$ Interest & $-$0.0024 & $-$0.0021 & $-$0.0027 & $-$0.0018 \\
& (0.004) & (0.001) & (0.003) & (0.001) \\
\midrule
$R^2$ & 0.013 & 0.019 & 0.014 & 0.022 \\
$N$ & 400 & 400 & 400 & 400 \\
\bottomrule
\end{tabular*}
\parbox{\textwidth}{\footnotesize\textit{Notes:} OLS regressions of price accuracy measures using bid-ask midpoint on treatment indicators. Standard errors in parentheses. * $p<0.10$, ** $p<0.05$, *** $p<0.01$.}
\end{table}

The results using the midpoint price are broadly consistent with those using the last trade price, though with smaller effect sizes and reduced statistical significance. The horizon effect remains significant for mean price accuracy ($p = 0.012$ for Price-Truth MAE) but is not significant for final-round prices ($p = 0.112$). The interaction terms are not statistically significant, though they have the expected negative sign. The weaker results with midpoint prices may reflect that the midpoint is less sensitive to trading activity than the last trade price.

\subsubsection{Mann-Whitney U Tests with Midpoint Price}

Tables \ref{tab:mwu_midpoint_h1}--\ref{tab:mwu_midpoint_h2_residual} report Mann-Whitney U tests for price accuracy measures using the bid-ask midpoint.

\begin{table}[htbp]
\centering
\caption{Mann-Whitney U Tests (Midpoint): H1 Long-Horizon Effect (Cell 2 vs. Cell 1)}
\label{tab:mwu_midpoint_h1}
\begin{tabular}{llrr}
\toprule
Outcome & Predicted Direction & $U$ Statistic & $p$-value \\
\midrule
Price-Truth MAE (Last) & Cell 2 $>$ Cell 1 & 3275.5 & 0.164 \\
Price-Truth MAE (Mean) & Cell 2 $>$ Cell 1 & 6093.5 & 0.004 \\
Price-Belief MAE (Last) & Cell 2 $>$ Cell 1 & 3302.0 & 0.142 \\
Price-Belief MAE (Mean) & Cell 2 $>$ Cell 1 & 6227.0 & 0.001 \\
\bottomrule
\end{tabular}
\parbox{\textwidth}{\footnotesize\textit{Notes:} One-sided Mann-Whitney U tests comparing Cell 2 to Cell 1 using bid-ask midpoint as the price measure.}
\end{table}

\begin{table}[htbp]
\centering
\caption{Mann-Whitney U Tests (Midpoint): H2 Direct Effect of Interest (Cell 4 vs. Cell 2)}
\label{tab:mwu_midpoint_h2_direct}
\begin{tabular}{llrr}
\toprule
Outcome & Predicted Direction & $U$ Statistic & $p$-value \\
\midrule
Price-Truth MAE (Last) & Cell 4 $<$ Cell 2 & 1972.0 & 0.053 \\
Price-Truth MAE (Mean) & Cell 4 $<$ Cell 2 & 3792.0 & 0.002 \\
Price-Belief MAE (Last) & Cell 4 $<$ Cell 2 & 1946.5 & 0.043 \\
Price-Belief MAE (Mean) & Cell 4 $<$ Cell 2 & 3622.0 & $<$0.001 \\
\bottomrule
\end{tabular}
\parbox{\textwidth}{\footnotesize\textit{Notes:} One-sided Mann-Whitney U tests comparing Cell 4 to Cell 2 using bid-ask midpoint as the price measure.}
\end{table}

\begin{table}[htbp]
\centering
\caption{Mann-Whitney U Tests (Midpoint): H2 Residual Horizon Effect (Cell 4 vs. Cell 3)}
\label{tab:mwu_midpoint_h2_residual}
\begin{tabular}{llrr}
\toprule
Outcome & Predicted Direction & $U$ Statistic & $p$-value \\
\midrule
Price-Truth MAE (Last) & Cell 4 $>$ Cell 3 & 2906.5 & 0.621 \\
Price-Truth MAE (Mean) & Cell 4 $>$ Cell 3 & 5185.0 & 0.326 \\
Price-Belief MAE (Last) & Cell 4 $>$ Cell 3 & 2705.0 & 0.848 \\
Price-Belief MAE (Mean) & Cell 4 $>$ Cell 3 & 5271.0 & 0.254 \\
\bottomrule
\end{tabular}
\parbox{\textwidth}{\footnotesize\textit{Notes:} One-sided Mann-Whitney U tests comparing Cell 4 to Cell 3 using bid-ask midpoint as the price measure.}
\end{table}

The Mann-Whitney results using the midpoint price show that the horizon effect (H1) is significant for mean price accuracy measures but not for final-round prices. The interest mitigation effect (H2) is significant for mean measures and marginally significant for final-round measures. Importantly, the residual horizon effect with interest is not significant for any measure ($p > 0.25$ in all cases), consistent with the finding that interest payments largely eliminate the horizon penalty.

\end{document}